\documentclass[preprint,12pt]{elsarticle}

\journal{Annals of Physics}

\usepackage{amsbsy}
\usepackage{amsfonts}
\usepackage{amsmath}
\usepackage{amssymb}
\usepackage{anyfontsize}
\usepackage{array}
\usepackage{bbding}
\usepackage{bbm}
\usepackage{bm}
\usepackage{color}
\usepackage{curves}
\usepackage{epsfig}
\usepackage{etoolbox}
\makeatletter
\AtBeginEnvironment{thebibliography}{%
   \clubpenalty10000
   \@clubpenalty \clubpenalty
   \widowpenalty10000}
\makeatother
\usepackage{fixmath}
\usepackage[T1]{fontenc}
\usepackage[margin=1in]{geometry}
\usepackage{graphicx}
\usepackage{multirow}
\usepackage{pdflscape}
\usepackage{placeins}
\usepackage[svgnames]{xcolor}
\usepackage{tikz}
\usetikzlibrary{svg.path}
\usepackage{relsize}
\usepackage{rotating}
\usepackage{scalerel}
\usepackage{txfonts}
\usepackage{ulem}

\definecolor{RoyalBlue}{cmyk}{1, 0.50, 0, 0}
\definecolor{ao(english)}{rgb}{0.0, 0.5, 0.0}
\definecolor{darkgreen}{RGB}{30,150,30}
\definecolor{darkblue}{RGB}{0,0,130}
\definecolor{darkred}{rgb}{.8,0,0}
\definecolor{orcidlogocol}{HTML}{A6CE39}

\DeclareMathAlphabet{\vecfont}{T1}{cmr}{bx}{it}
\DeclareMathAlphabet{\nfont}{T1}{cmr}{bx}{sl}
\DeclareMathAlphabet{\mathantt}{OT1}{antt}{li}{it}
\DeclareMathAlphabet{\mathpzc}{OT1}{pzc}{m}{it}

\newcommand{\negphantom}[1]{\ifmmode\settowidth{\dimen0}{$#1$}\else\settowidth{\dimen0}{#1}\fi\hspace*{-\dimen0}}
\newcommand{\Ket}[1]{\rule[-2.65pt]{0.65pt}{10.75pt}\hspace{0.25ex}#1\big>}
\newcommand{\Bra}[1]{\big<#1\hspace{0.25ex}\rule[-2.65pt]{0.65pt}{10.75pt}}
\renewcommand{\vec}[1]{\mathbold{#1}}
\newcommand{\nn}{\nonumber}
\renewcommand{\d}{\mathrm{d}}
\newcommand{\dd}{\mathrm{d}}

\newcommand{\bok}[3]{\left<\right.\hspace{-0.5ex}{#1}\left.\hspace{-0.5ex}\right|{#2}\left|\right.\hspace{-0.5ex}{#3}\left.\hspace{-0.5ex}\right>}
\newcommand{\Bok}[3]{\Bra{#1}\hspace{0.5ex}#2\hspace{0.4ex}\Ket{#3}}

\newcommand{\bk}[2]{\left<\right.\hspace{-0.5ex}{#1}\left.\hspace{-0.5ex}\right|{#2}\left.\hspace{-0.5ex}\right>}

\newcommand{\trace}{\mathrm{tr}}

\newcommand{\ignore}[1]{\relax}

\newcommand{\D}{\mathrm{d}}
\newcommand{\I}{\mathrm{i}}
\newcommand{\Exp}[1]{\mathrm{e}^{\mbox{\footnotesize$#1$}}}
\newcommand{\ket}[1]{{\left|{#1}\right\rangle}}
\newcommand{\bra}[1]{{\left\langle{#1}\right|}}
\newcommand{\braket}[2]{{\langle{#1}|{#2}\rangle}}
\newcommand{\expect}[1]{{\left\langle{#1}\right\rangle}}
\newcommand{\tr}[2][]{\mathrm{tr}_{#1}{\left(#2\right)}}

\newcommand{\half}{\frac{1}{2}}

\newcommand{\Min}[2][]{\mathop{\mathrm{Min}}_{#1}{\left\{#2\right\}}}
\newcommand{\Max}[2][]{\mathop{\mathrm{Max}}_{#1}{\left\{#2\right\}}}
\newcommand{\MIN}[2]{\underset{#1}{\mathrm{Min}}\left\{#2\right\}}
\newcommand{\vts}{\rule{0.1em}{0ex}}  
\newcommand{\column}[2][c]{{\left(\begin{array}{#1}#2\end{array}\right)}}

\newcommand{\dens}[1]{\boldsymbol{#1}}
\DeclareMathAlphabet{\cfont}{T1}{cmss}{sbc}{n}

\newcommand{\Varepsilon}{\mbox{\resizebox{0.7em}{!}{$\varepsilon$}}}

\renewcommand{\emph}{\textit}

\newcommand{\pqty}[1]{\left(#1\right)}
\newcommand{\pdv}[1]{\frac{\partial}{\partial #1}}
\newcommand{\Herm}[2]{\mathrm{H}_{#1} {\left({#2}\right)}}
\newcommand{\Ell}{\mathcal{L}}
\newcommand{\Int}[1][-5pt]{\int\limits_{\begin{picture}(16,3)(-8,-3)
		\put(0,0){\curve(-3,0,-8,0)\curve(3,0,8,0)}%
		\put(8,0){\curve(0,0,-1.5,1.5)\curve(0,0,-1.5,-1.5)}%
		\put(0,0){\arc(-3,0){180}}\put(0,0){\makebox(0,0){$\cdot$}}%
		\end{picture}}\hspace*{#1}}

\tikzset{
  orcidlogo/.pic={
    \fill[orcidlogocol] svg{M256,128c0,70.7-57.3,128-128,128C57.3,256,0,198.7,0,128C0,57.3,57.3,0,128,0C198.7,0,256,57.3,256,128z};
    \fill[white] svg{M86.3,186.2H70.9V79.1h15.4v48.4V186.2z}
                 svg{M108.9,79.1h41.6c39.6,0,57,28.3,57,53.6c0,27.5-21.5,53.6-56.8,53.6h-41.8V79.1z M124.3,172.4h24.5c34.9,0,42.9-26.5,42.9-39.7c0-21.5-13.7-39.7-43.7-39.7h-23.7V172.4z}
                 svg{M88.7,56.8c0,5.5-4.5,10.1-10.1,10.1c-5.6,0-10.1-4.6-10.1-10.1c0-5.6,4.5-10.1,10.1-10.1C84.2,46.7,88.7,51.3,88.7,56.8z};
  }
}
\newcommand\orcid[1]{\href{https://orcid.org/#1}{\mbox{\scalerel*{
\begin{tikzpicture}[yscale=-1,transform shape]
\pic{orcidlogo};
\end{tikzpicture}
}{|}}}}

\begin{document}

\begin{frontmatter}



\title{Single-particle-exact density functional theory}




\author[CQT]{Martin-Isbj\"orn~Trappe\corref{cor1}}
\ead{martin.trappe@quantumlah.org}
\cortext[cor1]{corresponding author}

\author[CQT]{Jun~Hao~Hue}
\ead{jhhue@nus.edu.sg}

\author[CQT]{Jonah~Huang~Zi~Chao}
\ead{e0309310@u.nus.edu}

\author[CQT]{Miko{\l}aj~Paraniak}
\ead{mparaniak@protonmail.com}

\author[LMU]{Djamila~Hiller}
\ead{d.hiller@physik.uni-muenchen.de}

\author[USZ,MPI]{Jerzy~Cios{\l}owski}
\ead{jerzy@wmf.univ.szczecin.pl}

\author[CQT,Beijing,DoP]{Berthold-Georg~Englert}
\ead{englert@u.nus.edu}

\affiliation[CQT]{organization={Centre for Quantum Technologies, National University of Singapore},
            addressline={3 Science Drive 2}, 
            postcode={Singapore 117543}, 
            country={Singapore}}

\affiliation[LMU]{organization={Fakult\"at f\"ur Physik, Ludwig-Maximilians-Universit\"at M\"unchen},
            addressline={Geschwister-Scholl-Platz 1}, 
            postcode={80539 M\"unchen}, 
            country={Germany}}

\affiliation[USZ]{organization={Institute of Physics, University of Szczecin},
            addressline={Wielkopolska 15}, 
            postcode={70-451 Szczecin}, 
            country={Poland}}

\affiliation[MPI]{organization={Max-Planck-Institut f\"ur Physik komplexer Systeme},
            addressline={N\"othnitzer Stra{\ss}e 38}, 
            postcode={01187 Dresden}, 
            country={Germany}}

\affiliation[Beijing]{organization={Key Laboratory of Advanced Optoelectronic Quantum Architecture and Measurement of Ministry of Education, School of Physics, Beijing Institute of Technology},
            postcode={Beijing 100081}, 
            country={China}}

\affiliation[DoP]{organization={Department of Physics, National University of Singapore},
            addressline={2 Science Drive 3}, 
            postcode={Singapore 117542}, 
            country={Singapore}}

\begin{abstract}
We introduce `single-particle-exact density functional theory' (1pEx-DFT), a novel density functional approach that represents all single-particle contributions to the energy with exact functionals. Here, we parameterize interaction energy functionals by utilizing two new schemes for constructing density matrices from `participation numbers' of the single-particle states of quantum many-body systems. These participation numbers play the role of the variational variables akin to the particle densities in standard orbital-free density functional theory. We minimize the total energies with the help of evolutionary algorithms and obtain ground-state energies that are typically accurate at the one-percent level for our proof-of-principle simulations that comprise interacting Fermi gases as well as the electronic structure of atoms and ions, with and without relativistic corrections. We thereby illustrate the ingredients and practical features of 1pEx-DFT and reveal its potential of becoming an accurate, scalable, and transferable technology for simulating mesoscopic quantum many-body systems.
\end{abstract}



\begin{keyword}

orbital-free density functional theory \sep Levy--Lieb constrained search \sep reduced density matrices \sep fermion gases \sep relativistic electronic structure \sep evolutionary algorithms



\end{keyword}

\end{frontmatter}



\newpage

\section{Introduction}

\begin{figure}[htb!]
\begin{center}
\includegraphics[width=0.75\linewidth]{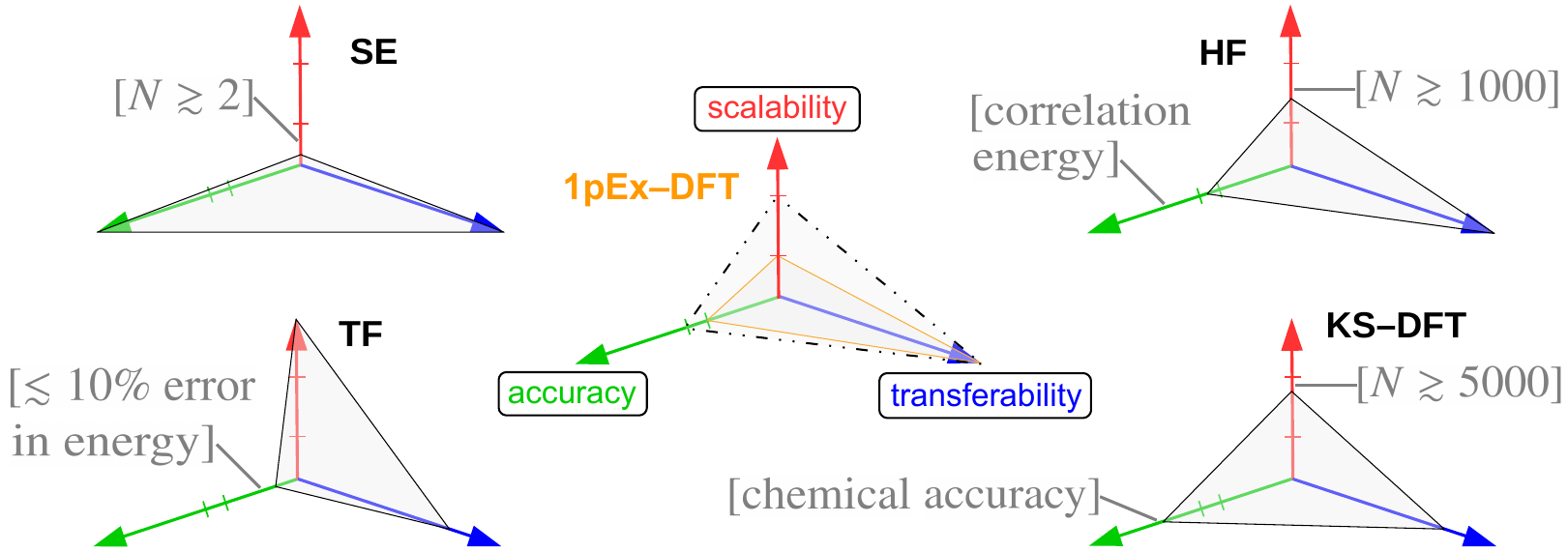}
\caption{\label{FigureTrinity}
The trade-off between accuracy, scalability, and transferability is inherent to all quantum many-body methods and also provides a means of classifying them, with the Schr\"{o}dinger equation (SE) at one end of this scale. Though orbital-free in spirit, our proposed `single-particle-exact-DFT' (1pEx-DFT) sits at the intersection of wave-function-based methods and density-based methods. While the current implementation of 1pEx-DFT (orange solid-line triangle) is completely transferable like HF, it is not yet as accurate and scalable as other established many-body methods. However, we discuss remedies that would let 1pEx-DFT complement orbital-free DFT as well as KS-DFT in significant ways (dash-dotted line). This is no small feat, but even if only partially successful, such a program could eventually supersede HF. We emphasize, however, that this enterprise is speculative at present. Square brackets indicate objectives that are \textit{typically} out of reach for the respective methods with common computational resources.}
\end{center}
\vspace{-100pt}
\end{figure}

The many variants of density functional theory (DFT) have been developed predominantly for calculating observables in position space---fueled by the decades-long success story of DFT applications in chemistry and materials science, see Refs.~\cite{Becke2014,Hasnip2014,Okun2023DFMPS} and references therein---alongside a mere handful of scholarly articles devoted to density functionals in momentum space \cite{Henderson1981,Cinal1993}, which target, for instance, Compton profiles \cite{Sakurai1995} or momentum distributions in ultracold quantum gases \cite{Hueck2018}. The ever increasing demand of high-quality solutions to specific quantum many-body problems from across scientific disciplines has been inciting DFT developers to refine the established methods and codes as well as to develop approaches that differ distinctly from prevailing ones in the hope of discovering a methodology that is superior at least for a subset of problems---typically by sacrificing one of the competing objectives of accuracy, scalability, and transferability (in the sense of universal applicability), see Fig.~\ref{FigureTrinity}. While the Schr\"{o}dinger equation is, by definition, entirely accurate and transferable, its numerical solution is typically limited to a few interacting particles. By neglecting inter-particle correlations beyond the exchange energy, Hartree--Fock (HF) theory offers much better scalability at the expense of accuracy, and quantum chemistry methods like the coupled cluster technique fall in between HF and the Schr\"{o}dinger equation. In contrast to these Hamiltonian-based approaches, the interaction energy for typical formulations of DFT like Thomas--Fermi-DFT (TF-DFT) and Kohn--Sham-DFT (KS-DFT) has to be constructed explicitly for any given interaction. This reduces transferability in practice. The TF kinetic energy functional offers unsurpassed scalability but is not accurate enough for predicting even the existence of molecular bonds. The most widely used KS-DFT often comes close to chemical accuracy, at the expense of transferability, but is typically limited to hundreds of particles.


\newpage

Historically, the search for alternatives to KS-DFT produced successful approaches like density matrix functional theory \cite{Levy1979,Levy1995,Savin1995,Goedecker1998,Cioslowski2000,Giesbertz2019}, which aims at chemical accuracy and transferability across electronic structure problems---and whose technical aspects are very much related to our agenda in the present article---as well as modern orbital-free DFT in all its variety, e.g., density-potential functional theory (DPFT) \cite{Englert1982,Trappe2016,Trappe2017,Chau2018,Englert2019articleEntry}, which aims at scalability toward large particle numbers and transferability across interactions and dimensionality. Still, these alternative methods are currently implemented in position space. But the Levy--Lieb constrained search \cite{Levy1979,Lieb1983}, which constitutes the fundamental justification of modern DFT, invites us to consider density functionals beyond the familiar configuration- and momentum-space representations. Here, we may hope to find new and promising many-body methods off the beaten path.

Reference~\cite{Englert2023DFMPS} delivers a unified view of these possibilities through a second-quantized perspective and proposes one particular DFT formulation based on `participation numbers' (referred to as `occupation numbers' in Ref.~\cite{Englert2023DFMPS}) of the eigenstates of the single-particle part of quantum many-body systems. This `single-particle-exact-DFT' (1pEx-DFT) presents a stark departure from the established formulations of DFT. Here, we introduce the details of the general 1pEx-DFT formalism and showcase applications to atomic gases and electronic structure that communicate the practical aspects of 1pEx-DFT. Although 1pEx-DFT is universally applicable, its current implementation may not capture beyond-HF correlations efficiently enough, see Ref.~\cite{Cioslowski2023}. While Hartree--Fock-level accuracy is usually considered insufficient for most chemistry applications, it is adequate for describing mesoscopic quantum gases \cite{Trappe2021b}, for which we believe 1pEx-DFT could prove particularly useful.

In Sec.~\ref{1pExDFT}, we derive the explicit formulation of the Levy--Lieb constrained search over the single-particle participation numbers (Sec.~\ref{GeneralFormalism}). For the proof-of-principle examples studied in this work, we parameterize the interaction energy in Dirac approximation \cite{Dirac1930} (or Hartree's and Fock's \cite{Hartree1928,Slater1930,Fock1930}, if you like): we omit correlations beyond the HF exchange energy (see also \ref{AppendixPerturbationTheory}) and develop two new constructions of one-body reduced density matrices from participation numbers (Sec.~\ref{InteractionTensorElementsDensityMatrices}); details are provided in \ref{AppendixApproximateRDMs}. In Sec.~\ref{EnergyMinimization} we discuss our explicit implementation of the Levy--Lieb constrained search. To minimize the total energy, we deploy stochastic evolutionary algorithms, specifically, particle swarm optimization \cite{Kennedy1995,Bonyadi2017,Tang2021} and a genetic algorithm \cite{Slowik2020}, see \ref{AppendixEA} for details. The required interaction tensor elements (known as two-electron integrals in the case of Coulomb interactions) are derived in \ref{AppendixInteractionTensorElements}. Our results on energies, participation numbers, and single-particle densities of Fermi gases and atomic systems in Sec.~\ref{Results} demonstrate the numerical feasibility of 1pEx-DFT. We benchmark our 1pEx-DFT predictions against HF results. In Sec.~\ref{Conclusions}, we discuss fundamental as well as technical challenges of 1pEx-DFT---regarding accuracy, scalability, and transferability---that ought to be overcome for realizing its prospective advantages over existing many-body methods.

\section{\label{1pExDFT}$\mathbf{1pEx}$-DFT}

In this section we derive the exact ground-state energy functional in terms of the single-particle participation numbers. Aiming at a first illustration of 1pEx-DFT, we also develop an explicit scheme that encodes the interaction energy in Dirac approximation and carries out the Levy--Lieb constrained search. We then show how to construct the required density matrices from an iterative algorithm and, alternatively, from a TF-inspired derivation in rotor phase space.

\subsection{\label{GeneralFormalism}General formalism}
We denote the position and momentum operators of the $j$th particle by
$\vec{R}_j$ and $\vec{P}_j$, respectively, and 
consider many-particle Hamilton operators of the generic form
\begin{align}\label{eq:A1}
  H_{\mathrm{mp}}=\sum_{j=1}^N\frac{1}{2m}\vec{P}_j^2
                   +\sum_{j=1}^NV_{\mathrm{ext}}\big(\vec{R}_j\big)
                   +\half\mathop{\sum_{j,k=1}^N}_{(j\neq k)}
                   V_{\mathrm{int}}\big(\vec{R}_j-\vec{R}_k\big)
                 =\sum_{j=1}^NH_{\mathrm{1p}}\big(\vec{P}_j,\vec{R}_j\big)
                     +H_{\mathrm{int}}\,,
\end{align}
with a sum of single-particle contributions and a sum over pairs for the
interaction contribution.
The one-particle Hamilton operator $H_{\mathrm{1p}}$ (the `core Hamiltonian') consists of the kinetic
energy of a particle with mass~$m$ and the potential energy in the external
potential $V_{\mathrm{ext}}$ that traps the particles,
\begin{align}
  \label{eq:A2}
  H_{\mathrm{1p}}(\vec{P},\vec{R})=
  \frac{1}{2m}\vec{P}^2+V_{\mathrm{ext}}(\vec{R})\,.
\end{align}
The system is composed of $N$ identical spin-$\half$ particles, with symmetric pair interaction energy ${V_{\mathrm{int}}\big(\vec{R}_j-\vec{R}_k\big)=V_{\mathrm{int}}\big(\vec{R}_k-\vec{R}_j\big)}$.
More complicated situations are thinkable, such as an external electric or
magnetic field and a corresponding change in $H_{\mathrm{1p}}$, a
spin-dependent interaction, or alternative dispersion relations.

The eigenstates of $H_{\mathrm{1p}}$ constitute a complete orthonormal set of
\textit{orbital} states (the `1pEx-basis'),
\begin{align}
  \label{eq:A3}
  H_{\mathrm{1p}}(\vec{P},\vec{R})\ket{a}
  =\ket{a}E_{a}\,,\quad
  \braket{a}{b}=\delta_{a,b}\,,\quad
  \sum_{a}\ket{a}\bra{a}=1\,,
\end{align}
here written for discrete quantum numbers $a$.
More generally, there could also be a continuous part of the spectrum
(scattering states) with corresponding modifications of the orthonormality
and completeness relations. For the sake of notational simplicity, we will
employ the conventions of Eq.~(\ref{eq:A3}) while keeping in mind that there could be
adjustments if $H_{\mathrm{1p}}$ has scattering states, which are commonly disregarded in DFT calculations.

The $H_{\mathrm{1p}}$-induced single-particle density
${\dens{n}=(n_{1},n_{2},\dots)}$ is the list of `participation numbers'
\begin{align}
  \label{eq:A4}
  n_{a}=\expect{\sum_{j=1}^N\Bigl(\ket{a}\bra{a}\Bigr)_j}\,,
\end{align}
evaluated in the applicable many-particle state. The $j$th term $\expect{\bigl(\ket{a}\bra{a}\bigr)_j}$
is the probability of finding the $j$th particle in the $a$th orbital
state. For simplicity we restrict ourselves in the following to unpolarized systems with even $N$. Hence,
\begin{align}\label{OccNumConstraint1}
0\leq n_{a}\leq2
\end{align}
as there can be at most two
spin-$\half$ particles in the same orbital state.
The completeness of the orbital states for each particle ensures that the sum
of all $n_{a}$ is equal to the particle count,
\begin{align}
  \label{OccNumConstraint2}
  \sum_{a}n_{a}=\expect{\sum_{j=1}^N1}=N\,.
\end{align}
Since we have 
\begin{align}
H_{\mathrm{1p}}(\vec{P},\vec{R})=\sum_{a}\ket{a}E_{a}\bra{a}\,, 
\end{align}
it follows that
\begin{align}
  \label{eq:A7}
  E_{\mathrm{1p}}=\expect{\sum_{j=1}^NH_{\mathrm{1p}}(\vec{P}_j,\vec{R}_j)}
  =\sum_{a}n_{a}E_{a}\,,
\end{align}
the sum of the single-particle energies $E_{a}$ weighted by the
participation numbers $n_{a}$. Equation~(\ref{eq:A7}) will typically entail that a large part of the total energy is treated exactly and that characteristic consequences of the external potential, such as the electron cusp in atoms, are automatically built in through the 1pEx-basis.

The ground-state energy $E_{\mathrm{gs}}$ of the many-particle Hamilton
operator in Eq.~(\ref{eq:A1}) is the minimum of all expectation values of
$H_{\mathrm{mp}}$,
\begin{align}
  \label{eq:A8}
  E_{\mathrm{gs}}
  =\Min[\rho^{\ }_{\mathrm{mp}}]
  {\tr{\rho_{\mathrm{mp}}H_{\mathrm{mp}}}\rule{0pt}{10pt}}\,,
\end{align}
where all $N$-particle statistical operators $\rho^{\ }_{\mathrm{mp}}$
participate in the competition. While it would be sufficient to consider
pure-state $\rho_{\mathrm{mp}}$s, there is no need for this restriction. 
In the spirit of the Levy--Lieb constrained search in standard DFT, we
regard this minimization as a two-step process: first we minimize over the
$\rho_{\mathrm{mp}}$s that yield a prescribed single-particle density
$\dens{n}$, then we minimize over all permissible $\dens{n}\vts$s, 
\begin{align}
  \label{eq:A9}
  E_{\mathrm{gs}}
  =\Min[\dens{n}]{ E_{\mathrm{mp}}[\dens{n}]\rule{0pt}{10pt}}
\end{align}
with
\begin{align}
  \label{eq:A9b}
    E_{\mathrm{mp}}[\dens{n}]=\Min[\rho^{\ }_{\mathrm{mp}}\to\dens{n}]
   {\tr{\rho_{\mathrm{mp}}H_{\mathrm{mp}}}\rule{0pt}{10pt}}\,,
\end{align}
where, as a consequence of Eq.~(\ref{eq:A7}),
\begin{align}
  \label{eq:A10}
  E_{\mathrm{mp}}[\dens{n}]=\sum_{a}n_{a}E_{a}+
  \Min[\rho^{\ }_{\mathrm{mp}}\to\dens{n}]
  {\tr{\rho_{\mathrm{mp}}H_{\mathrm{int}}}\rule{0pt}{10pt}}=\sum_{a}n_{a}E_{a}+ E_{\mathrm{int}}[\dens{n}]
\end{align}
is the \textit{single-particle-exact density functional}.

Indeed, the contribution from the sum of single-particle energies is exact
and simple in $ E_{\mathrm{mp}}[\dens{n}]$, whereas the contribution from the
pair interactions requires a minimization over all permissible
$\rho_{\mathrm{mp}}$s.
Usually, we cannot find this minimum and have to be content
with a suitable approximation. We will work with one such approximation in Sec.~\ref{ExOnly1pEXDFT} below.

For a given $\rho_{\mathrm{mp}}$, we have the single-particle-orbital density
matrix in position space
\begin{align}
  \label{eq:B1}
  n^{(1)}(\vec{r};\vec{r}')
  =\tr{\rho_{\mathrm{mp}}\sum_{j=1}^N\Bigl(\Ket{\vec{r}'}\Bra{\vec{r}}\Bigr)_j}\,,
\end{align}
here expressed in orbital states and normalized to the particle count ${\int(\D\vec{r})\,  n^{(1)}(\vec{r};\vec{r})=N}$. We also have the two-particle orbital-density matrix
\begin{align}
  \label{eq:B3}
  n^{(2)}(\vec{r}_1^{\ },\vec{r}_2^{\ };\vec{r}_1',\vec{r}_2')
  =\half\tr{\rho_{\mathrm{mp}}\mathop{\sum_{j,k=1}^N}_{(j\neq k)}
    \Bigl(\Ket{\vec{r}_1'}\Bra{\vec{r}_1^{\ }}\Bigr)_j
    \otimes\Bigl(\Ket{\vec{r}_2'}\Bra{\vec{r}_2^{\ }}\Bigr)_k}\,,
\end{align}
which is symmetric under particle exchange, ${n^{(2)}(\vec{r}_1^{\ },\vec{r}_2^{\ };\vec{r}_1',\vec{r}_2')=n^{(2)}(\vec{r}_2^{\ },\vec{r}_1^{\ };\vec{r}_2',\vec{r}_1')}$, and related to the single-particle density by  
\begin{align}
  \label{eq:B5}
  \int(\D\vec{r}_2^{\ })\,
  n^{(2)}(\vec{r},\vec{r}_2^{\ };\vec{r}',\vec{r}_2^{\ })
  =\half(N-1)\, n^{(1)}(\vec{r};\vec{r}')\,.
\end{align}
It follows that the full trace of $n^{(2)}$ is the count of pairs, 
\begin{align}
\int(\D\vec{r})\,(\D\vec{r}')\,  n^{(2)}(\vec{r},\vec{r}';\vec{r},\vec{r}')=\half N(N-1)\,. 
\end{align}
As indicated these are \textit{orbital} densities, that is, the spin variables
are traced out.
This is fine under the given circumstances as we have no spin dependence in
$H_{\mathrm{1p}}$ and $H_{\mathrm{int}}$.

The constraint `$\rho_{\mathrm{mp}}\to\dens{n}$' in Eqs.~(\ref{eq:A9})--(\ref{eq:A10}) means
\begin{align}
  \label{eq:B7}
  n_{a}^{\ }=\tr{\rho_{\mathrm{mp}}
    \sum_{j=1}^N\Bigl(\ket{a}\bra{a}\Bigr)_j}
  =  \int(\D\vec{r})\,(\D\vec{r}')\,\psi_{a}(\vec{r})^*
         \,n^{(1)}(\vec{r};\vec{r}')\,\psi_{a}(\vec{r}')\,,
\end{align}
where $\psi_{a}(\vec{r})=\braket{\vec{r}}{a}$ is the wave function of the
$a$th eigenstate of $H_{\mathrm{1p}}$.
We introduce the effective (orbital) single-particle statistical operator $\rho$ through
\begin{align}
  \label{eq:B8}
  &n^{(1)}(\vec{r};\vec{r}')
  =\Bok{\vec{r}}{\rho}{\vec{r}'}
  \quad\mbox{or}\quad
  \rho=
  \int(\D\vec{r})\,(\D\vec{r}')\,\ket{\vec{r}}
     \,n^{(1)}(\vec{r};\vec{r}')\,\langle\vec{r}'|\,,
     \quad\mbox{with}\quad
  \rho\geq0\,,\ 
  \tr{ \rho}=N\,,
\end{align}
and then Eq.~(\ref{eq:B7}) reads
\begin{align}
  \label{eq:B9}
  n_{a}^{\ }=\bok{a}{\rho}{a}\,.
\end{align}
Since $n_{a}\leq2$, the eigenvalues of $\rho$ cannot
exceed $2$, ${\rho\leq2}$. It follows that the rank of $\rho$ is $\half N$ or larger.

Since
\begin{align}
  \label{eq:C1}
  {\tr{\rho_{\mathrm{mp}}H_{\mathrm{int}}}\rule{0pt}{10pt}}
  =\int(\D\vec{r})\,(\D\vec{r}')\,V_{\mathrm{int}}(\vec{r}-\vec{r'})
   \,n^{(2)}(\vec{r},\vec{r}';\vec{r},\vec{r}')\,,
\end{align}
the minimization in Eq.~(\ref{eq:A10}) requires us to consider all $n^{(2)}$s that
yield, via Eq.~(\ref{eq:B5}) and Eq.~(\ref{eq:B8}), a $\rho$ that obeys
Eq.~(\ref{eq:B9}) for the given $n_{a}^{\ }$s.
We do not have a generic parameterization of $n^{(2)}$ that is suitable for this
purpose and, therefore, resort to approximations, that is, rather than
minimizing over \textit{all} permissible $n^{(2)}$s, we minimize over a smaller
set and then work with the resulting approximation for
$E_{\mathrm{int}}[\dens{n}]$.

\subsection{\label{ExOnly1pEXDFT}Implementation of an exchange-only 1pEx-DFT}

Dirac's approximation (that is, the Hartree--Fock approximation \cite{Hartree1928,Slater1930,Fock1930}) for the two-particle density matrix in terms of the
single-particle density matrix \cite{Dirac1930},
\begin{align}
  \label{eq:Dirac}
    n^{(2)}(\vec{r}_1^{\ },\vec{r}_2^{\ };\vec{r}_1',\vec{r}_2')
    =\half n^{(1)}(\vec{r}_1^{\ };\vec{r}_1')
           \,n^{(1)}(\vec{r}_2^{\ };\vec{r}_2')
    -\frac{1}{4} n^{(1)}(\vec{r}_1^{\ };\vec{r}_2')
                 \,n^{(1)}(\vec{r}_2^{\ };\vec{r}_1')\,,
\end{align}
has a very good track record, and we employ it.
This is exact for pure-state $\rho_{\mathrm{mp}}$s that are single Slater
determinants and yield single-particle density matrices with a ${2\times2}$
spin matrix that is a multiple of the identity; the particle number $N$ has to
be even for that, hardly a restriction as we are mostly interested in
situations with very many particles.
For Dirac's $n^{(2)}$, the integral in Eq.~(\ref{eq:B5}) states
\begin{align}\label{eq:C3}
  \half N\, n^{(1)}(\vec{r};\vec{r}')
  -\frac{1}{4}\int(\D\vec{r}_2^{\ })\, n^{(1)}(\vec{r};\vec{r}_2^{\ })
  \,n^{(1)}(\vec{r}_2^{\ };\vec{r}')
  =\half (N-1)\, n^{(1)}(\vec{r};\vec{r}')\,,
\end{align}
That is,
\begin{align}
  \label{eq:C4}
  \int(\D\vec{r}_2^{\ })\, n^{(1)}(\vec{r};\vec{r}_2^{\ })
  \,n^{(1)}(\vec{r}_2^{\ };\vec{r}')=2 n^{(1)}(\vec{r};\vec{r}')
\end{align}
or
\begin{align}
  \label{eq:C5}
  \rho^2=2\rho\,,
\end{align}
when expressed as a property of the statistical operator.
Accordingly, in Dirac's approximation,
$\rho$ has $\half N$ eigenvalues equal to
$2$ and all other eigenvalues are zero.
Any two such $\rho$s are related to each other by a unitary
transformation, and the constraints in Eq.~(\ref{eq:B9}) require that the subspaces
specified by an $n_{a}^{\ }$ value are invariant under the unitary
transformation. In other words, the minimization in Eq.~(\ref{eq:A10}) needs to be carried out over all unitary transformations that leave the diagonal elements
$\bok{a}{\rho}{a}$ of the single-particle statistical operator 
\begin{align}
  \label{eq:D1}
  \rho=\sum_{a,b}
  \ket{a}\,\varrho^{\ }_{ab}\,\bra{b}
\end{align}
unchanged. Here, with the emphasis on the dependence on $\dens{n}$, it is natural to use the
1pEx-basis $\{\ket{a}\}$ for the parameterization of $\rho$.

One family of suitable unitary transformations multiplies each $\ket{a}$ by a phase
factor
\begin{align}
  \label{eq:C6}
  \ket{a}\to\ket{a}\,\Exp{\I\phi_{a}}\,.
\end{align}
Taking into account all admissible transformations that preserve individually the diagonal elements in ${2\times2}$ sectors of $\varrho$, that is, more general transformations than those in Eq.~(\ref{eq:C6}), we find no further improvement in the ground-state energies of Eq.~(\ref{eq:A9}), see \ref{AppendixApproximateRDMs} for details. For the purpose of this work we are thus content with optimizing the phases~$\phi_{a}$ in Eq.~(\ref{eq:C6}) and will attend elsewhere to the expansion of the search space toward all permissible density matrices. We want to emphasize, however, that we deem the use of density matrices an auxiliary measure that is straightforward but comes with an inflated number of parameterization variables. Here, these are the phases~$\phi_{a}$ on top of the participation numbers $n_a$. Ultimately, we hope to construct more efficient functionals $E_{\mathrm{int}}[\dens{n}]$ without introducing an excessive number of parameterization variables (not to be confused with free, adjustable parameters).

Upon using Dirac's approximation, viz., Eq.~(\ref{eq:Dirac}), in Eq.~(\ref{eq:C1}) and recalling Eq.~(\ref{eq:B8}), we
have 
\begin{align}
  \label{eq:C7}
    {\tr{\rho_{\mathrm{mp}}H_{\mathrm{int}}}\rule{0pt}{10pt}}
  &=\half\int(\D\vec{r})\,(\D\vec{r}')\,V_{\mathrm{int}}(\vec{r}-\vec{r'})
       \,\left(n^{(1)}(\vec{r};\vec{r})\,n^{(1)}(\vec{r}';\vec{r}')
      -\frac{1}{2}
     n^{(1)}(\vec{r};\vec{r}')\,n^{(1)}(\vec{r}';\vec{r})\right)
     \nn\\
  &=\half
      \int(\D\vec{r})\,(\D\vec{r}')\,V_{\mathrm{int}}(\vec{r}-\vec{r'})
      \,\left(\bok{\vec{r}}{\rho}{\vec{r}}
      \,\bok{\vec{r}'}{\rho}{\vec{r}'}
      -\frac{1}{2}
      \bok{\vec{r}}{\rho}{\vec{r}'}
      \,\bok{\vec{r}'}{\rho}{\vec{r}}\right)\,,\qquad\quad
\end{align}
where we exploit the Fourier integral for 
\begin{align}
V_{\mathrm{int}}(\vec{r}-\vec{r'})=\int\frac{(\D\vec{k})}{(2\pi)^{D}}\,  u(\vec{k})\,\Exp{\I\vec{k}\cdot\vec{r}}\Exp{-\I\vec{k}\cdot\vec{r}'}
\end{align}
for $D$ spatial dimensions to factorize the $\vec{r}$ and $\vec{r'}$ dependence; since
$V_{\mathrm{int}}(\vec{r})$ is real and even, so is~$u(\vec{k})$. Then, with 
\begin{align}
\int(\D\vec{r})\,\Exp{\I\vec{k}\cdot\vec{r}} \bok{\vec{r}}{\rho}{\vec{r}}=\tr{\rho\,\Exp{\I\vec{k}\cdot\vec{R}}}\,,
\end{align}
for example, we arrive at
\begin{align}
  \label{eq:C10}
  {\tr{\rho_{\mathrm{mp}}H_{\mathrm{int}}}\rule{0pt}{10pt}}
  =\half \int\frac{(\D\vec{k})}{(2\pi)^{D}}\, u(\vec{k})\,\left(
  \;\biggl|\tr{\rho\,\Exp{\I\vec{k}\cdot\vec{R}}}
     \biggr|^2-\half\tr{\rho\,\Exp{\I\vec{k}\cdot\vec{R}}
     \,\rho\,\Exp{-\I\vec{k}\cdot\vec{R}}}\right)\,.
     \rule{30pt}{0pt}
\end{align}
In Eq.~(\ref{eq:A10}), we minimize this over all $\rho$ that are
permitted by Eqs.~(\ref{eq:B9}) and (\ref{eq:C5}).

With Eq.~(\ref{eq:D1}) we have
\begin{align}
  \label{eq:D2}
  \tr{\rho\,\Exp{\I\vec{k}\cdot\vec{R}}}
  &=\sum_{a,b}\,\varrho^{\ }_{ab}
      \Bok{b}{\Exp{\I\vec{k}\cdot\vec{R}}}{a}\\
  \intertext{and}
  \tr{\rho\,\Exp{\I\vec{k}\cdot\vec{R}}
  \,\rho\,\Exp{-\I\vec{k}\cdot\vec{R}}}
  &=\sum_{a,b}\sum_{c,d}
      \varrho^{\ }_{ab}
      \Bok{b}{\Exp{\I\vec{k}\cdot\vec{R}}}{c}
                    \,\varrho^{\ }_{cd}
      \Bok{d}{\Exp{-\I\vec{k}\cdot\vec{R}}}{a}\,.
\end{align}
Then,
\begin{align}
  \label{eq:D3}
  \tr{\rho_{\mathrm{mp}}H_{\mathrm{int}}}
  =\half \sum_{a,b}\sum_{c,d}
  \varrho^{\ }_{ab}
  \,\mathcal{H}^{\ }_{ab,cd}
  \,\varrho^{\ }_{cd}
\end{align}
with $\mathcal{H}^{\ }_{ab,cd}=I_{abcd}-\frac12 I_{adcb}$, where
\begin{align}
I_{abcd}&=\int(\d\vec r)(\d\vec r')\,V_{\mathrm{int}}(\vec r-\vec r')\,\psi_a(\vec r)\psi_b^*(\vec r)\psi_c(\vec r')\psi_d^*(\vec r')\nn\\
&=\frac{1}{(2 \pi)^D}\int \d\vec k\, u(\vec k)\, \Bok{a}{\Exp{\I \vec{k}\cdot\vec{R}}}{b}\, \Bok{c}{\Exp{-\I \vec k\cdot\vec{R}}}{d}\label{Tabcd}
\end{align}
is a set of interaction-energy tensor elements (commonly known as two-electron integrals in the case of coulombic electron--electron interactions), which are completely
determined by the interaction potential $V_{\mathrm{int}}$ and the
orbital eigenstates of the single-particle Hamilton operator
$H_{\mathrm{1p}}$.

We have thus explicitly formulated the constrained search in Eq.~(\ref{eq:A9b}) over $\rho_{\mathrm{mp}}$ as a constrained search over density matrices $\varrho$ in the 1pEx-basis, where $\varrho$ derives from $\rho_{\mathrm{mp}}$ through Eqs.~(\ref{eq:D1}), (\ref{eq:B8}), and (\ref{eq:B1}). In any particular application, we need
\begin{align}\label{eq:D5}\fboxsep=5pt%
\fbox{\begin{tabular}{r@{\ }p{250pt}}
     (i) & {\raggedright{}to find the values of the
$\mathcal{H}^{\ }_{ab,cd}$ coefficients;}
\\  (ii) & {\raggedright{}to write down one $\varrho^{(0)}$ matrix
                   for $\rho$ with\newline
           ${\varrho^{(0)}_{aa}=n^{\ }_{a}}$
           and ${\left(\varrho^{(0)}\right)^2=2\,\varrho^{(0)}}$;}
\\ and (iii) & {\raggedright{}to minimize the right-hand-side
                       of Eq.~(\ref{eq:D3}) with
${\varrho^{\ }_{ab}=\Exp{\I\phi_{a}}\varrho^{(0)}_{ab}\Exp{-\I\phi_{b}}}$
over all phases ${\boldsymbol{\phi}=(\phi_{1},\phi_{2},\dots)}$.}                    
      \end{tabular}}
\end{align}    
The resulting density functional $E_{\mathrm{mp}}[\dens{n}]$ can then be minimized over all permissible $\dens{n}$. In practice, we minimize over $\dens{n}$ and $\boldsymbol{\phi}$ simultaneously---the concrete implementation of Eq.~(\ref{eq:D5}) is described in the next Secs.~\ref{InteractionTensorElementsDensityMatrices} and \ref{EnergyMinimization}.

\subsection{\label{InteractionTensorElementsDensityMatrices} Interaction tensor elements and density matrices}

For any practical computation, the program outlined in Eq.~(\ref{eq:D5}) comes with two approximations. First, the Dirac approximation of the interaction energy in terms of the one-body density matrix $\varrho$---we leave for future work the inclusion of correlation energy expressed in terms of $\varrho$. Second, the incomplete set of matrices $\varrho$ that enters the competition in Eq.~(\ref{eq:A9b}), which can be improved by (i) going beyond the transformations with phase factors $\Exp{\I\phi_{a}}$ (toward all admissible $\varrho$, that is, toward full HF), (ii) considering different seed matrices $\varrho^{(0)}$, and (iii) increasing the dimension $L$ of $\varrho$, which has to be finite in practice. Additional approximations may stem from the interaction tensor elements $I_{abcd}$ in Eq.~(\ref{Tabcd}) if the orbital eigenstates and energies of $H_{\mathrm{1p}}$ can only be obtained numerically or approximately. For the systems considered in the present work, however, these ingredients are analytical or quasi-exact, see \ref{AppendixInteractionTensorElements} for details. Hence, for the purpose of this article, we are content with approximating the ground-state energy assembled from Eqs.~(\ref{eq:A9})--(\ref{eq:A10}) and Eq.~(\ref{eq:D3}) by
\begin{align}
  \label{E}
  E_{\mathrm{gs}}
  \approx\MIN{\boldsymbol{\Theta},\boldsymbol{\phi}}{\sum_{a=1}^L (1+\cos\theta_{a})\,E_{a}+\half \sum_{a,b,c,d=1}^L \varrho^{(0)}_{ab}\,\varrho^{(0)}_{cd}\,\left(I_{abcd}-\half I_{adcb}\right)\,\mathrm{e}^{\I\,(\phi_{a}-\phi_{b}+\phi_{c}-\phi_{d})}},
\end{align}
where we incorporate the Dirac approximation of Eq.~(\ref{eq:Dirac}) and take into account the finite number $L$ of 1pEx-basis states. We write ${n_a=1+\cos\theta_{a}}$ with angles ${\Theta=(\theta_1,\dots,\theta_L)}$, such that we satisfy ${0\leq n_a\leq2}$ automatically and only need to enforce the particle count of Eq.~(\ref{OccNumConstraint2}). Finally, in general we have ${I_{abcd}=I_{badc}^*}$ and ${\varrho_{ab}=\varrho_{ba}^*}$, while for the case studies in this work, ${V_{\mathrm{int}}(\vec r-\vec r')=V_{\mathrm{int}}(\vec r'-\vec r)}$, and both $I_{abcd}$ and $\varrho_{ab}^{(0)}$ are real. Therefore, we may replace the phase factor in Eq.~(\ref{E}) by ${\cos(\phi_{a}-\phi_{b}+\phi_{c}-\phi_{d})}$.

We obtain a seed matrix ${\varrho^{(0)}=\varrho^{\mathrm{it}}}$ with admissible entries by iteratively transforming the diagonal matrix with diagonal $(2,2,\dots,2,0,0,\dots,0)$ and trace $N$. In each step of this matrix mixer algorithm introduced in Ref.~\cite{ParaniakPhdThesis2022} we apply a unitary operation that produces one target diagonal element ${\varrho^{\mathrm{it}}_{aa}=n_a}$. That is, after at most ${L-1}$ steps, we arrive at a proper density matrix (obeying ${\varrho^2=2\varrho}$, cf.~Eq.~(\ref{eq:C5})) with prescribed diagonal elements and non-zero off-diagonal elements, see \ref{AppendixApproximateRDMs} for details.

If we take into account enough 1pEx-basis states, then any discrepancies between 1pEx-DFT results (obtained using $\varrho^{\mathrm{it}}$) and those from HF reported in this work stem from the incomplete parameterization of the density matrices. In such cases, improvements of the 1pEx-DFT energies toward the HF energies have to come from transformations beyond those that individually preserve the diagonal elements in ${2\times2}$ sectors of the density matrices. It is an open problem to determine for which systems such more general transformations would make our current 1pEx-DFT implementation equivalent to HF.

As an alternative for 1D systems, we use the approximate density matrix
\begin{align}\label{rhotf}
\varrho^{\mathrm{tf}}_{ab}=\frac{g\sin\left((a-b)\,\sigma\right)}{\pi(a-b)}\,,
\end{align}
with degeneracy factor $g$ (for unpolarized spin-$\half$ fermions, ${g=2}$) and $\cot\sigma=\half\Big[\cot\left(\frac{\pi}{2}n_a\right)+\cot\left(\frac{\pi}{2}n_b\right)\Big]$. We obtain Eq.~(\ref{rhotf}) from an approximate Wigner function in rotor phase space, inspired by the classical-phase-space argument that yields the TF-approximated spatial density matrix ${\varrho_{\mathrm{TF}}(x;x')}$. That is, ${\varrho^{(0)}=\varrho^{(0)}(\dens{n})}$ in Eq.~(\ref{E}) stands for either $\varrho^{\mathrm{it}}$ or $\varrho^{\mathrm{tf}}$, which are derived in \ref{AppendixApproximateRDMs}.

\subsection{\label{EnergyMinimization}Energy minimization}

The angles $\theta_a$ and phases $\phi_a$ form the search space for the minimization of $E_{\mathrm{gs}}$ in Eq.~(\ref{E}) and are only constrained by Eq.~(\ref{OccNumConstraint2}), which spans an $(L-1)$-dimensional hypersurface within the $L$-dimensional space of participation numbers ~$\dens{n}$. Suitable choices of global optimizers for constrained non-convex functions in high-dimensional spaces are problem-dependent. We opt for stochastic algorithms, especially evolutionary algorithms, which are efficient in optimizing our constrained high-dimensional non-convex energy functions for two reasons: first, stochastic optimization allows us to project (improper) randomly requested vectors $\dens{n}$ onto the constraining hypersurface before we evaluate the energy. Second, evolutionary algorithms can be easily tuned to escape even deep local optima efficiently and can optimize highly deceptive objective functions such as Eq.~(\ref{E}).

We consulted two evolutionary algorithms: a particle swarm optimization (PSO) \cite{Kennedy1995,Bonyadi2017,Tang2021} and a genetic-algorithm optimizer (GAO) \cite{Slowik2020}, see \ref{AppendixEA} for the details of our implementations, which outperformed---on average, for the test cases we considered---coupled simulated annealers, adapted from Ref.~\cite{deSouza2010}. As an alternative to genuinely stochastic optimizers, we used multi-start linearly-constrained conjugate gradient optimization, based on the C++ implementation of the $\mbox{ALGLIB}$ project at www.alglib.net.
In our test cases, all four aforementioned algorithms produced virtually the same numerical values for the ground-state participation numbers $\dens{n}$. However, PSO proved superior among our implementations of optimizers regarding the convergence speed toward the optima of our case studies; all results of optimizations reported in this work are obtained with PSO, unless explicitly stated otherwise.

\section{\label{Results}Results on energies, participation numbers, and spatial densities}

We applied the 1pEx-DFT program entailed in Eq.~(\ref{eq:D5}) and Secs.~\ref{InteractionTensorElementsDensityMatrices} and \ref{EnergyMinimization} to harmonically confined fermions in 1D, subjected to contact interaction (Sec.~\ref{Sec1DContact}) and harmonic interaction (Sec.~\ref{Sec1DHarmonic}). Moreover, in Sec.~\ref{RelativisticAtoms} we extract the electronic structure of atoms and ions from 1pEx-DFT, both with and without relativistic corrections from spin-free exact two-component Dirac theory \cite{Kutzelnigg2005,Liu2009,Cheng2011,Cunha2022}. With our choice of the Dirac approximation for the interaction energy $E_{\mathrm{int}}[\dens{n}]$, HF theory is the natural choice for benchmarking our simulations: Like in HF theory, the Dirac approximation is the only physical approximation that enters our current 1pEx-DFT implementation, such that both methods must produce the same results if the energy minimization covers all admissible density matrices, which is the case when minimizing Eq.~(\ref{E}) with ${\varrho^{(0)}=\varrho^{\mathrm{it}}}$ for ${N=2}$, cf. Tables~I and II below. Future implementations of 1pEx-DFT along the same lines of the present work may exceed HF accuracy if we use approximations of the interaction energy that include at least some correlation but can still be expressed in terms of the one-body density matrix. While HF theory is reasonably accurate and transferable, it is orbital-based and, hence, of limited scalability. Therefore, keeping in mind 1pEx-DFT applications to mesoscopic systems, we also compared with a variant of orbital-free DPFT that is scalable and transferable. While this semiclassical method tends to become (relatively) accurate only for larger particle numbers, it comes with an established track record across systems and disciplines \cite{Englert1982,Trappe2016,Trappe2017,Chau2018,Trappe2019,Englert2019articleEntry,Trappe2021b,Trappe2023DFMPS,TrappeWittManzhosXXXX,Trappe2023NatComm}.

\subsection{\label{Sec1DContact}Harmonically trapped contact-interacting fermions}
We begin with $N$ spin-$\half$ fermions in 1D harmonic confinement and pair interaction energy
\begin{align}\label{VintContact}
V_{\mathrm{int}}(x-x')=\mathpzc{c}\,\delta(x-x')\,.
\end{align}
The corresponding interaction tensor elements $I_{abcd}$ are derived in \ref{AppendixContactInteraction1D}. All numerical values in Secs.~\ref{Sec1DContact} and \ref{Sec1DHarmonic} are given in harmonic oscillator units of energy $\hbar\omega$ and length $\sqrt{\hbar/(m\omega)}$, with particle mass $m$ and oscillator frequency $\omega$. That is, for example, an interaction strength ${\mathpzc{c}=20}$ in Eq.~(\ref{VintContact}) is implicit for ${\mathpzc{c}=20\,\hbar\omega\sqrt{\hbar/(m\omega)}}$.

As expected, the energies $E_{\mathrm{1pEx}}^{\mathrm{tf}}$ based on the TF inspired $\varrho^{\mathrm{tf}}$ are hardly accurate for small $N$, but Table~\ref{1DcontactTable} suggests that they become relatively more accurate with increasing $N$. The energies $E_{\mathrm{1pEx}}^{\mathrm{it}}$ are more accurate for the cases considered here and approach the HF energies to within one or two percent. By comparison, the DPFT energies are quite accurate when using the exact HF interaction energy ${E_{\mathrm{int}}^{\mathrm{HF}}=(\mathpzc{c}/4)\int\d x\,\big(n(x)\big)^2}$, see \ref{AppendixApproximateRDMs}. However, unlike HF theory, the approximate DPFT energy functionals are not variational and do not guarantee to deliver upper bounds to $E_{\mathrm{gs}}$---here, the TF density $n_{\mathrm{TF}}$ incidentally delivers an even lower energy than its quantum-corrected successor $n_{3'}$---and systematic improvements beyond $n_{3'}$ can incur high computational cost \cite{Chau2018,Trappe2023DFMPS}.

If we assume that the number $L$ of single-particle levels required to converge the energy in Eq.~(\ref{E}) scales like the particle number $N$, then the computational cost of the \textit{current} implementation of 1pEx-DFT scales like $\mathcal O\big(N^4\big)$, compared with the generic scaling of $\mathcal O\big(N^3\big)$ for HF and KS--DFT. The cost of the DPFT densities $n_{\mathrm{TF}}$ and $n_{3'}$ scales like $\mathcal O\big(N\big)$ and $\mathcal O\big(N^2\big)$, respectively. Of course, these scaling behaviors are really informative only in the limit of large $N$, and the same scaling does not imply the same cost in practice, with HF and KS--DFT presenting a point in case. Furthermore, the actually realized computational cost much depends on the system. For example, we found the energies $E^{\mathrm{it}}_{\mathrm{1pEx}}$ in Table~I with 1 CPU-hour (for ${c=1}$, ${N=4}$, ${L=20}$), 16 CPU-hours (for ${c=1}$, ${N=10}$, ${L=20}$), and 59 CPU-days (for ${c=20}$, ${N=10}$, ${L=30}$), respectively. We list these timings only for completeness, bearing in mind that our focus here is to illustrate the concepts behind a new approach, not to optimize code and CPU time.

\begin{center}
\begin{table}[ht]
\setlength\extrarowheight{0.4em}
\setlength\tabcolsep{0.31em}
\begin{tabular}{r|r|ccc|cc|cc}
$\mathpzc{c}$ & $N$ & $E^{\mathrm{it}}_{\mathrm{1pEx}}$ & $E_{\mathrm{HF}}$ & $\big(E_{\mathrm{1pEx}}^{\mathrm{it}}-E_{\mathrm{HF}}\big)/E_{\mathrm{HF}}$ & $E_{\mathrm{1pEx}}^{\mathrm{tf}}$ & $\big(E_{\mathrm{1pEx}}^{\mathrm{tf}}-E_{\mathrm{HF}}\big)/E_{\mathrm{HF}}$ & $E_{\mathrm{DPFT}}(n_{\mathrm{TF}})$ & $E_{\mathrm{DPFT}}(n_{3'})$ \\
\hline\hline
\multirow{4}{*}{1} & 2   & 1.3790 & 1.3790 & 0.00\% & 1.3243 & -3.97\% & 1.3648 & 1.4526 \\
                   & 4   & 5.0642 & 5.0590 & 0.10\% & 4.9773 & -1.62\% & 5.0451 & 5.2420 \\
                   & 10  & 29.218 & 29.193 & 0.09\% & 29.053 & -0.48\% & 29.181 & 29.353 \\
                   & 20  & 111.97 & 111.91 & 0.05\% & 111.75 & -0.14\% & 111.90 & 112.12 \\
\hline
\multirow{4}{*}{20} & 2   & 5.9695 & 5.9695 & 0.00\% & 6.3862 & 6.98\% & 5.9214 & 6.0391 \\
                    & 4   & 19.416 & 19.083 & 1.75\%  & 19.999 & 4.80\% & 19.031 & 19.211 \\
                    & 10  & 90.572 & 90.031 & 0.60\%  & 93.486 & 3.84\% & 89.974 & 90.155 \\
                    & 20  & 298.60 & 294.75 & 1.31\%  & 304.45 & 3.29\% & 294.69 & 294.89 \\
\end{tabular}
\caption{\label{1DcontactTable}Comparison of $E_{\mathrm{1pEx}}$ and HF energies $E_{\mathrm{HF}}$ for $N$ spin-$\half$ fermions in a 1D harmonic trap at contact-interaction strengths ${\mathpzc{c}=1}$ and ${\mathpzc{c}=20}$. Since both 1pEx-DFT and HF are exact for noninteracting systems, it is not surprising that the two methods agree better for weaker interactions. In contrast to $\varrho^{\mathrm{it}}$, the TF-inspired density matrix $\varrho^{\mathrm{tf}}$ breaks the variational character of Eq.~(\ref{E}) and can therefore yield energies smaller than $E_{\mathrm{HF}}$, as we see here for ${\mathpzc{c}=1}$. Also the density approximations $n_{\mathrm{TF}}$ and $n_{3'}$ breach the variational property of the general DPFT framework. Incidentally, energies below $E_{\mathrm{HF}}$ materialize here with the TF approximation, but not with the generically superior approximation associated with the semiclassical density $n_{3'}$. Efficient implementations of $n_{3'}$ from Ref.~\cite{Chau2018} are provided in Refs.~\cite{Trappe2021b,TrappeWittManzhosXXXX}. Here and in Table \ref{1DharmonicTable} below, we report HF energies calculated from the Roothaan equations for closed-shell systems in the 1pEx-basis, with the level-shifting procedure of Ref.~\cite{Saunders1973} to aid convergence in the case of strong interactions.}
\end{table}
\end{center}

For ${N=2}$, the density matrix $\varrho^{\mathrm{it}}$ obtained from the iterative algorithm of Sec.~\ref{InteractionTensorElementsDensityMatrices} is identical to the HF density matrix $\varrho^{\mathrm{HF}}$---up to phases that are optimized anyway during the energy minimization---see \ref{AppendixApproximateRDMs}, such that 1pEx-DFT recovers HF exactly. However, HF is inadequate for modeling this particular system: with the exact ground-state wave function $\Psi$ for two spin-$\half$ fermions \cite{Busch+3:98, Viana-Gomes+1:11} and 
\begin{align}
n^{(1)} (x;x') = \bok{x}{\rho}{x'} = 2 \int \d x_2\, \Psi(x,x_2)\, \Psi(x',x_2)^*\,, 
\end{align}
we calculate the exact density matrix in the 1pEx-basis, ${\varrho_{ab}=\int \dd{x} \dd{x'}\,\bk{a}{x}\bok{x}{\rho}{x'}\bk{x'}{b}}$, and obtain ${E_{\mathrm{gs}}\approx1.92}$ for the true ground-state energy at ${\mathpzc{c}=20}$, which deviates from $E_{\mathrm{HF}}$ by a factor of three---in other words, the correlation energy dominates the total energy---and explains the large discrepancies between the exact and the HF/1pEx participation numbers shown in Fig.~\ref{1Dcontact_OccNum_2} in \ref{AppendixApproximateRDMs}. 

Figure~\ref{1Dcontact_OccNum_10} shows the converged participation numbers for ${N=10}$ obtained with PSO and GAO, respectively. We also present spatial densities extracted from the converged density matrix $\hat{\varrho}^{\mathrm{it}}_{ab}=\Exp{\I\hat{\phi}_{a}}\hat{\varrho}^{(0)}_{ab}\Exp{-\I\hat{\phi}_{b}}$, which comprises both the eventually successful seed matrix $\hat{\varrho}^{(0)}$ and the phases $\boldsymbol{\phi}$ that are optimal for $\hat{\varrho}^{(0)}$; see \ref{AppendixApproximateRDMs} for details. The participation numbers provide intuitively accessible information, since we usually have painless access to (and a solid understanding of) the noninteracting basis states (the 1pEx-basis)---we may imagine how unintuitive chemistry would be if we could not refer to the noninteracting hydrogenic energy levels 1s, 2s, 2p, etc., when talking about the real (interacting) electrons in atoms. For noninteracting $N$-particle systems the lowest $N/2$ levels of the 1pEx-basis are fully occupied (with participation numbers ${n_{1\le a\le N/2}=2}$), and all other levels are unoccupied (${n_{a>N/2}=0}$), such that the deviations from this distribution of participation numbers in the case of interacting systems are a measure of the interaction strength. The participation numbers can also reveal nontrivial structures like the irrelevance of the odd-parity states in Fig.~\ref{1Dcontact_OccNum_2} in \ref{AppendixApproximateRDMs} (if $\rho^{\mathrm{it}}$ were taken to be the truth).

\begin{figure}[htb!]
\begin{center}
\includegraphics[width=0.65\linewidth]{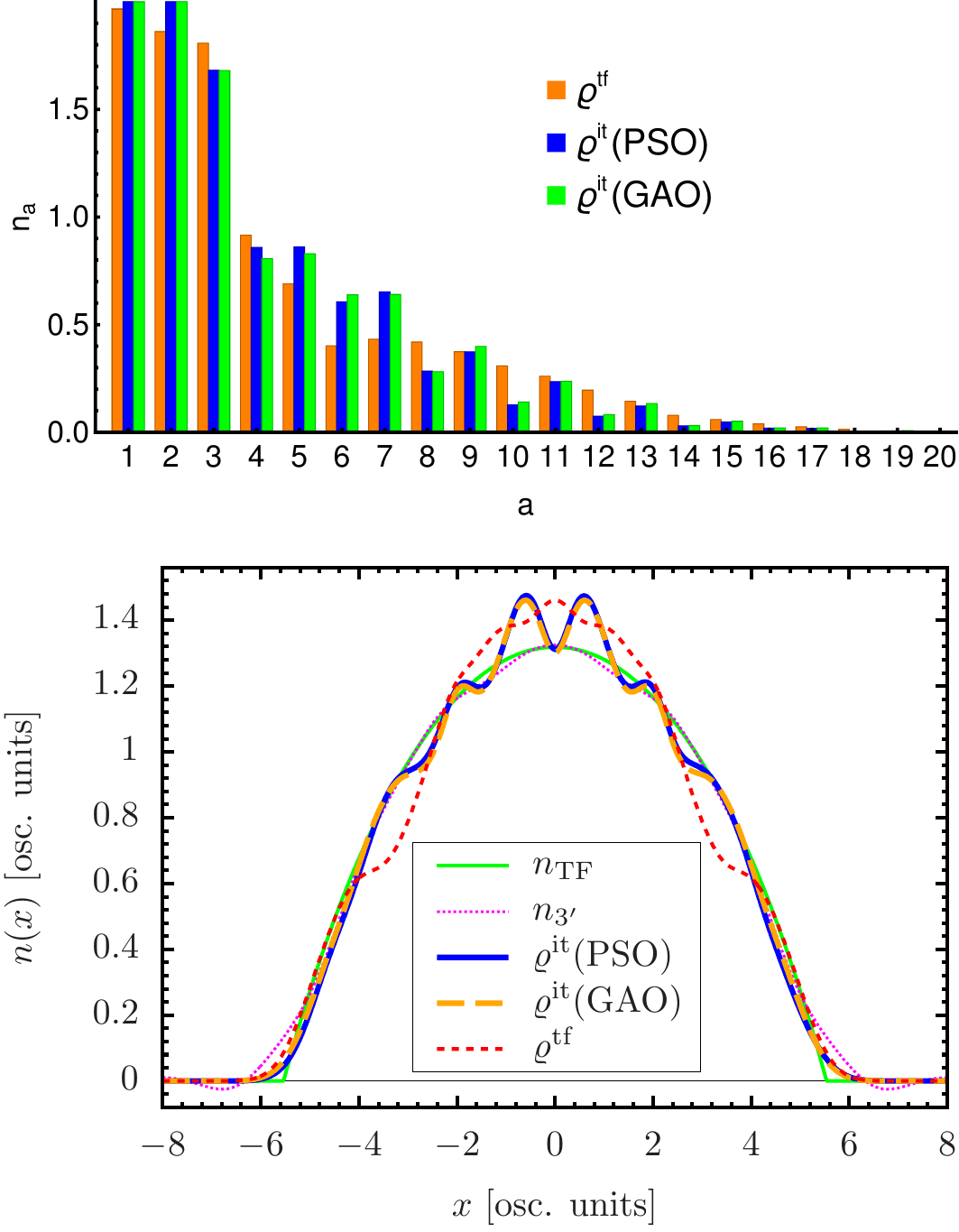}
\caption{\label{1Dcontact_OccNum_10} Participation numbers $n_a$ for single-particle levels ${a=1,\dots,20}$ (top) and spatial densities of ${N=10}$ spin-$\half$ fermions in a 1D harmonic trap at contact-interaction strength ${\mathpzc{c}=20}$ (bottom). The participation numbers obtained from PSO (and GAO, as a cross-check of our numerics) follow the same trend as those that originate in $\varrho^{\mathrm{tf}}$. We compute the spatial densities labeled `$\varrho^{\mathrm{it/tf}}$' from the converged density matrices, i.e., the seed matrices $\varrho^{(0)}$ combined with the optimized phases $\boldsymbol{\phi}$, and get the characteristic quantum-mechanical oscillations in the center of the trap, which are difficult to obtain with semiclassical orbital-free DFT methods like the DPFT implementation based on Refs.~\cite{Trappe2016,Trappe2017,Chau2018,Trappe2021b}, which yields, for example, the TF density $n_{\mathrm{TF}}$ and the quantum-corrected density $n_{3'}$ shown here.}
\end{center}
\end{figure}

\FloatBarrier

\subsection{\label{Sec1DHarmonic}Harmonically trapped fermions with harmonic interaction}

Next, we consider harmonically trapped fermions in 1D with the pair energy
\begin{align}\label{Vintharmonic}
V_{\mathrm{int}}(x_j-x_k)=\frac{m}{2N}(\Omega^2-\omega^2)\,(x_j-x_k)^2=\frac{\alpha-1}{2N}(x_j-x_k)^2\,,
\end{align}
where ${\alpha=\Omega^2/\omega^2}$ defines a dimensionless interaction strength. Equation~(\ref{Vintharmonic}) is expressed in the oscillator units $\hbar\omega$ and $\sqrt{\hbar/(m\omega)}$ of the noninteracting system (${\alpha=1}$). In \ref{AppendixHarmonicInteraction1D} we derive the interaction tensor elements for Eq.~(\ref{Vintharmonic}) via Eq.~(\ref{Tabcd}).

In Table~\ref{1DharmonicTable}, we benchmark the energies predicted by 1pEx-DFT for ${N=2}$ and ${N=20}$ with ${\alpha=3/2}$ against HF results and compare with the exact energies as well as with DPFT energies. The energies labeled `direct' include only the direct (Hartree) part of the interaction energy, i.e., ${\mathcal{H}^{\ }_{ab,cd}=I_{abcd}}$. Then, the exact energy is given by ${E_{\mathrm d}^{(\alpha)}=\sqrt{\alpha}N^2/4}$ for even $N$, see Eq.~(\ref{Ealpha}) in \ref{AppendixHarmonicInteraction1D}---the energy of an effectively noninteracting harmonic oscillator with frequency $\omega\sqrt{\alpha}$. We also note that Eq.~(\ref{Ealpha}) is a density functional and can be directly employed in DFT variants such as DPFT---for example, in TF approximation, which is known to deliver the exact energy for the noninteracting harmonic oscillator.

\begin{center}
\begin{table}[ht]
\setlength\extrarowheight{0.4em}
\setlength\tabcolsep{0.3em}
\begin{tabular}{c|r|c|ccr|cr|cc}
 & N & $E_{\mathrm d}^{(\alpha)}\;\;$ & $E^{\mathrm{it}}_{\mathrm{1pEx}}$ & $E_{\mathrm{HF}}$ & $\frac{E_{\mathrm{1pEx}}^{\mathrm{it}}}{E_{\mathrm{HF}}}-1$ & $E_{\mathrm{1pEx}}^{\mathrm{tf}}$ & $\frac{E_{\mathrm{1pEx}}^{\mathrm{tf}}}{E_{\mathrm{HF}}}-1$ & $E_{\mathrm{DPFT}}(n_{\mathrm{TF}})$ & $E_{\mathrm{DPFT}}(n_{3'})$ \\[0.25em]
\hline\hline
\multirow{2}{*}{\centering direct}             & 2   & 1.22474 & 1.22474 & 1.22474 & 0.00\% & 1.13746 & -7.13\% & 1.22474 & 1.51236 \\
                                    & 20  & 122.474 & 123.865 & 122.474 & 1.14\% & 124.156 & 1.37\% & 122.474 & 122.752 \\                                  
\hline
\multirow{2}{*}{\parbox{4em}{\vspace{1.0ex}\centering direct \&\\ exchange}} & 2   & --- & 1.11803 & 1.11803 & 0.00\% & 0.99835 & -10.7\% & --- & --- \\
                                                                  & 20  & --- & 123.605 & 122.368 & 1.01\% & 123.907 & 1.26\% & --- & --- \\
\end{tabular}
\caption{\label{1DharmonicTable}Like in Table~\ref{1DcontactTable}, we compare the optimized energies $E_{\mathrm{1pEx}}$ with the self-consistent energies $E_{\mathrm{HF}}$ for $N$ spin-$\half$ fermions in a 1D harmonic trap, but here for a harmonic interaction of strength ${\alpha=3/2}$. The exact energies for ${N=2}$ and ${N=20}$ at ${\alpha=3/2}$ are $1.11237$ and $122.362$, respectively \cite{BenavidesRiveros2014}. The direct (direct \& exchange) contributions to the interaction energy comprise about $20\%$ ($10\%$) of the total energy.}
\end{table}
\end{center}

\FloatBarrier

\subsection{\label{RelativisticAtoms}(Non-)relativistic atoms and ions}

We have emphasized transferability across quantum-mechanical systems as one advantageous feature that sets 1pEx-DFT apart from other DFT variants. While we believe that simulations of ultracold quantum gases will dominate the applications of 1pEx-DFT, single atoms and ions comprise an important point of departure for any novel quantum many-body method whose scope includes, in principle, atomic systems from molecules to nanoparticles to materials. As our last case study we therefore calculate the electronic structure of atoms and ions. We also extract binding energies of highly charged ions by using the (numerical) eigenfunctions of a relativistic core Hamiltonian. The latter is an example of the most general situation for the usage of 1pEx-DFT, where eigenstates and eigenenergies of the core Hamiltonian are only available in numerical form.

The Coulomb interaction energy for a pair of electrons at positions $\vec r$ and $\vec r'$ is $V_{\mathrm{int}}(\vec r\hspace{-0.07ex}-\hspace{-0.07ex}\vec r')=1/|\vec r\hspace{-0.07ex}-\hspace{-0.07ex}\vec r'|$. Accordingly, all numerical values in Sec.~\ref{RelativisticAtoms} are given in units of Hartree (Ha) and Bohr radius ($a_0$). The electrons are subjected to the external potential ${V_{\mathrm{ext}}(\vec r) = -Z/|\vec r|}$ of the nucleus with nuclear charge $Z$ at the origin of the spatial coordinate system, which makes the hydrogenic states our 1pEx-basis; details are provided in \ref{AppendixCoulombInteraction3D}.

In Fig.~\ref{RelDev_1pEx_HF_NIST} we benchmark the total binding energies of atoms and ions. We obtained the energies in Fig.~\ref{RelDev_1pEx_HF_NIST} by composing the 1pEx-basis solely of the hydrogenic bound states, see \ref{AppendixNonrelativisticAtoms}. When constrained to these states, HF indeed delivers the 1pEx energies for the two-electron systems reported in Fig.~\ref{RelDev_1pEx_HF_NIST}. However, the scattering states associated with the continuous part of the spectrum of the nuclear Coulomb potential need to be taken into account to recover the exact HF energies in the case of atomic systems with two electrons. In fact, with the scattering states completing the 1pEx-basis, 1pEx-DFT results should gain accuracy for electronic structure calculations in general; we leave this potentially fruitful enterprise for future work. In its current implementation with (i) the Dirac approximation and (ii) the incomplete search over density matrices, 1pEx-DFT yields energies that are accurate at the $1\mbox{--}2\%$~level when compared with HF, which itself provides the same level of accuracy when compared with the NIST Atomic Spectra Database \cite{NIST2021articleEntry}. As expected, the accuracy of $E_{\mathrm{1pEx}}$ improves as we increase the number $L$ of states of the 1pEx-basis that enter the competition in Eq.~(\ref{E}); as an illustration, we report binding energies for carbon with ${L=7}$, ${L=15}$, and ${L=31}$. Furthermore, $E_{\mathrm{1pEx}}$ becomes relatively more accurate as the ion-electron interaction intensifies. Then, deviations of the participation numbers $\dens{n}$ from the Fermi--Dirac distribution incur a penalty that increases, for example, along the helium isoelectronic sequence. In other words, the contributions of $H_{\mathrm{1p}}$ in highly charged ions dominate over the interaction energy, such that both the single Slater-determinant of HF theory and the exact single-particle treatment of 1pEx-DFT provide an increasingly accurate description, if relativistic effects are included when called for. This observation is also in line with our results for the carbon-like Xe$^{48+}$ and the neon-like Xe$^{44+}$. Also the diminishing influence of the scattering states on the binding energies along the helium isoelectronic series helps align our 1pEx energies with the HF benchmark in Fig.~\ref{RelDev_1pEx_HF_NIST}, see \ref{AppendixNonrelativisticAtoms} for details. By design, however, $E_{\mathrm{HF}}$ is a lower bound to the currently implemented $E_{\mathrm{1pEx}}$---which is also true when accounting for relativistic effects. They become more important with increasing nuclear charge; see Ref.~\cite{Fischer2016} for a review on the theory of complex atoms. Since 1pEx-DFT handles the nuclear singularity exactly for any $Z$ by adopting the 1pEx-basis, 1pEx-DFT can be directly applied to large-$Z$ atoms and ions---provided that we supply the relativistic hydrogenic states as 1pEx-basis. These states are known analytically from four-component Dirac theory \cite{Bethe1957}, but for simplicity we will stay within the confines of the nonrelativistic algebra that underpins the specific 1pEx-framework laid out in Sec.~\ref{GeneralFormalism}; see Ref.~\cite{Englert2023DFMPS} for the more general perspective on DFT from a second-quantized point of view. Since electron--positron pair production is irrelevant for chemistry, we can fully take into account the relativistic corrections using the Schr\"odinger-like exact two-component quasi-relativistic method (X2C) \cite{Kutzelnigg2005,Liu2009}. In fact, its spin-free approximation (sf-X2C) is accurate enough for our purposes, see \ref{AppendixRelativisticAtoms} for details. 

We use the sf-X2C Hamiltonian for calculating the relativistic HF energies that provide the benchmarks for our 1pEx(sf-X2C) energies in Fig.~\ref{RelDev_1pEx_HF_NIST}. The sf-X2C Hamiltonian also determines our relativistic 1pEx-basis and the associated interaction tensor elements; see \ref{AppendixRelativisticAtoms} for an outline of our numerical procedures. This application of 1pEx-DFT is an example of the generic situation, in which the eigenstates of $H_{\mathrm{1p}}$ can only be determined numerically.

\begin{figure}[htb!]
\begin{center}
\includegraphics[width=0.85\linewidth]{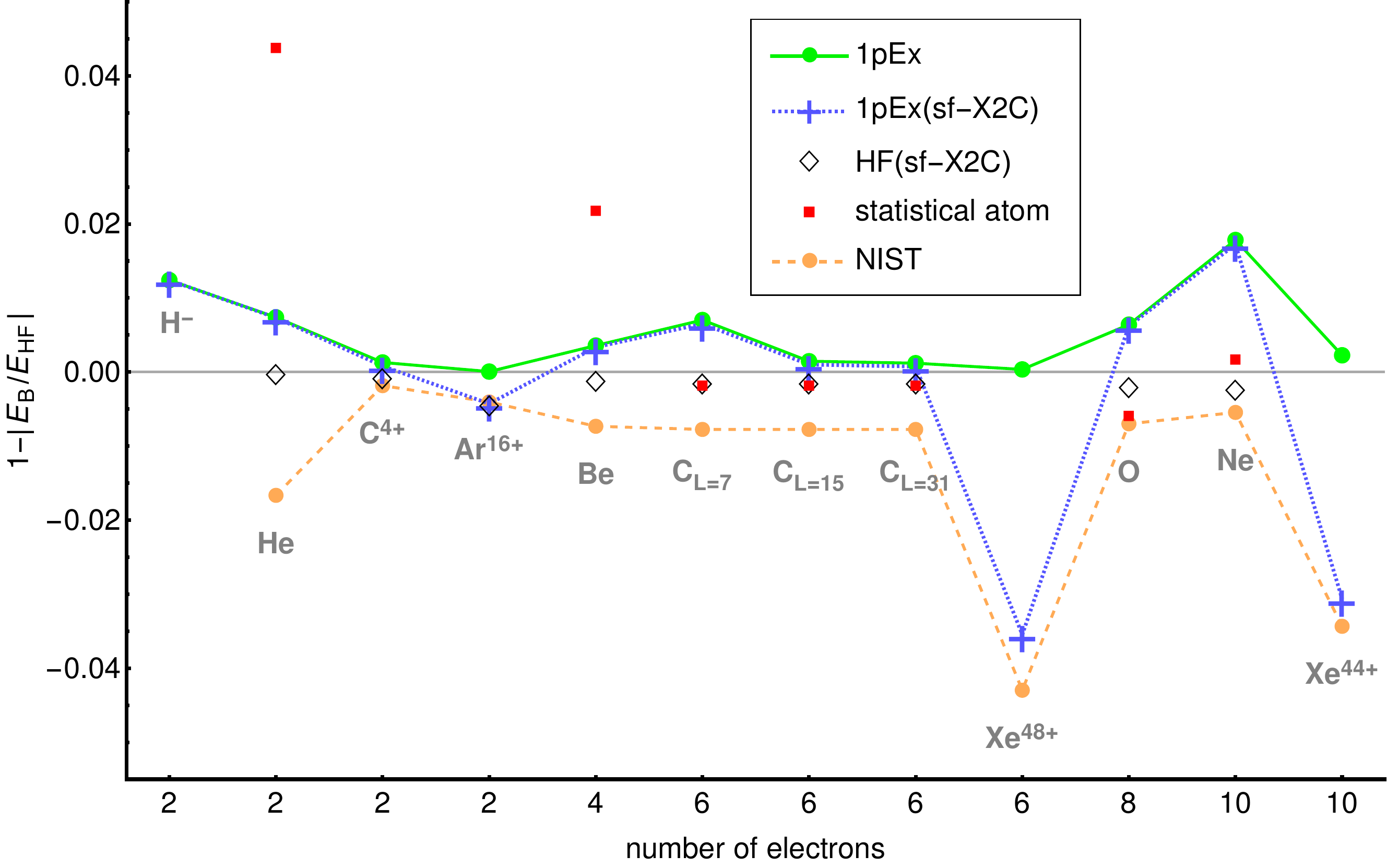}
\caption{\label{RelDev_1pEx_HF_NIST}
Total binding energies $E_{\mathrm{B}}$ of atoms and ions relative to the nonrelativistic HF energies $E_{\mathrm{HF}}$. The latter are an upper bound to the exact nonrelativistic energies and a lower bound to (our current implementation of) the nonrelativistic $E_{\mathrm{1pEx}}$. Smaller ordinates correspond to lower total ground-state energies (viz., negative binding energies). The connecting lines guide the eye. The absolute values of the energies displayed here are listed in Table~\ref{BindingEnergiesTable} in \ref{AppendixRelativisticAtoms}.}
\end{center}
\end{figure} 

Since Fig.~\ref{RelDev_1pEx_HF_NIST} displays relative deviations to the nonrelativistic $E_{\mathrm{HF}}$, the relativistic effects we extract from 1pEx(sf-X2C) are visible (to the eye) only for the highly charged ions. We also compare our results with a (nonrelativistic) approximation for the binding energy of neutral atoms: the `statistical atom' denotes a semiclassical approximation that includes the Scott correction as well as quantum corrections upon the TF approximation and becomes relatively more accurate as the atomic number increases \cite{Scott1952,Berge1988articleEntry,Englert2019articleEntry}. The binding energies calculated with 1pEx(sf-X2C) tend to approach the NIST data for highly charged ions, where relativistic energy corrections can dominate over correlation energy and other effects (such as QED effects) that are not taken into account here. In fact, both 1pEx(sf-X2C) and HF(sf-X2C) deliver essentially the exact reference value for helium-like Ar$^{16+}$. This contrasts with the cases of neutral helium and the hydrogen anion. For the latter, HF underestimates the experimental binding energy by more than 7\% \cite{Lykke1991}.

\FloatBarrier

\section{\label{Conclusions}Discussion and conclusions}

We introduced 1pEx-DFT, a novel generic quantum many-body method in the spirit of orbital-free DFT, where the energy functional depends on the particle density (in whichever representation), viz., the diagonal of the density matrix. This contrasts with density matrix functional theory, where the single-particle part is also exact but depends on the entire density matrix. 1pEx-DFT is as transferable as other explicitly Hamiltonian-based methods, for example, HF theory or the Schr\"{o}dinger equation itself. Here, we gave first illustrations of the broad scope of 1pEx-DFT by simulating interacting Fermi gases in one-dimensional confinement and by extracting relativistic corrections in the electronic structure of atoms and ions.

The proof-of-principle implementation of 1pEx-DFT in this work can reach HF accuracy by design, although full equivalence with HF in all cases will require an extension of the currently implemented constrained search toward all one-body density matrices compatible with the HF approximation. And since our analysis in \ref{AppendixPerturbationTheory}, see also Ref.~\cite{Cioslowski2023}, indicates that the evaluation of cumulant-based corrections to the HF approximation in the 1pEx-basis is costly, the accuracy of 1pEx-DFT could often be limited in practice to that of HF. This restriction is, however, minor for an important class of systems that we think should be targeted by 1pEx-DFT, namely ultracold atomic gases of mesoscopic size, for which the cost even of HF calculations is usually prohibitive. Figure~\ref{FigureTrinity} illustrates our take on the trinity of transferability, accuracy, and scalability of 1pEx-DFT, and we argue that the latter poses the primary challenge for 1pEx-DFT in becoming a complementing approach to today's workhorses among quantum many-body methods. In the following we propose several routes to attacking this issue of scalability.

The figure of merit for judging the efficacy of traditional optimizers for minimizing the energy in Eq.~(\ref{E}) is the number of function (or gradient) evaluations required to reach a targeted threshold for the function value. As a rule, this number grows sharply with the dimension $D$ of the search space. In addition, the computational cost of a single evaluation of Eq.~(\ref{E}), where ${D=2L}$, scales like $\mathcal O\big(D^4\big)$, such that our currently implemented objective-function-based optimizers, including PSO, are too costly for systems that require us to consider hundreds of single-particle levels. As an alternative to PSO, we may employ stochastic gradient descent (SGD) or its refinements \cite{Kingma2017}, which forfeit objective function evaluations and can escape local optima by descending along approximate gradients. The prototypical application of SGD is the minimization of functions that are sums like Eq.~(\ref{E}), such that the computational cost for a single gradient evaluation can be reduced from $\mathcal O\big(D^4\big)$ to $\mathcal O\big(1\big)$---at the expense of a potentially large number of descent steps, but with the promise of the same efficiency gains that also enable large-scale machine learning applications based on automatic differentiation. In particular, SGD on graphical processing units could prove promising since the interaction tensor elements $I_{abcd}$ of Eq.~(\ref{Tabcd}) cannot all be stored in memory and have to be (re-)calculated during runtime anyway if, say, ${D\gtrsim1000}$. This would also circumvent another caveat of 1pEx-DFT, namely that $I_{abcd}$ has to be recalculated when the external potential changes, for example, during the search for the ground-state geometry of molecules. A third and somewhat counterintuitive type of optimizer disregards both gradients \textit{and} function values: `novelty search' contrasts with traditional optimizers as it escapes local optima by guiding the optimizer into unexplored regions of the search space---a heuristic that can optimize highly deceptive objective functions \cite{Lehman2008}. We acknowledge that no strategy can guarantee finding the global optimum $g$ of a black-box function and that $g$ can only be found if the optimizer happens to arrive---ultimately by sheer luck, but hopefully accelerated via tried-and-tested heuristics---in the optimizer-specific attractor region of $g$. Hence, we can only hypothesize that combining a pool of optimizers will give us the means to minimize Eq.~(\ref{E}) across physical systems even in high-dimensional (${D\gtrsim1000}$) spaces. We also leave for future work the embedding of a suitable optimizer into a divide-and-conquer strategy for large-scale global optimization \cite{LaTorre2015,Glorieux2015,Sun2018}.

Maybe the greatest potential for scalability lies in explicit closed-form functionals $E_{\mathrm{int}}[\dens{n}]$ with computational cost ${\sim\mathcal O\big(D\big)}$, similar to $E_{\mathrm{1p}}$ in Eq.~(\ref{eq:A7}), which could make all the optimization strategies mentioned above feasible for systems with thousands of particles. As the most important quest in this regard we deem the search for a TF-type approximate functional that works at least for trapped atomic gases. Furthermore, when approaching a continuum of single-particle levels as $N$ increases, it may be possible to approximate the four-dimensional sum of Eq.~(\ref{eq:D3}) by interaction-specific integrals that can be evaluated more efficiently than the sum. Independent of the functional form of the interaction energy or choice of optimizer, we may also fully occupy the low-lying levels and optimize in the remaining `active space' of partially occupied levels---we already implicitly impose vanishing participation numbers of levels beyond $L$, just like any other method that imposes energy cutoffs. This reduces the optimization cost, especially for weakly interacting systems and, for example, for heavier atoms. Such a `frozen-core approximation' of 1pEx-DFT does not require ad-hoc assumptions or fits to data and is thus a natural ab-initio alternative to the use of pseudopotentials that approximate the Coulomb potential in traditional DFT methods. Of course, 1pEx-DFT calculations could also benefit directly from the use of pseudopotentials, if the implied uncertainties are tolerable, but the appeal of 1pEx-DFT comes in particular from the exact and unproblematic treatment of external potentials. For example, 1pEx-DFT naturally accounts for (and supersedes) the Scott correction \cite{Scott1952,Englert2019articleEntry}, the leading correction to the Thomas--Fermi model of atoms with singular nuclear potential.

We are confident that the list of aforementioned issues---and their remedies---is by no means complete. And all our suggestions for increasing the efficiency of 1pEx-DFT are uncharted territory at present. But, taking the cue from the history of the quantum many-body problem in general and the history of DFT in particular, we may hope to resolve over time many of the technical challenges that initially accompany a novel method like 1pEx-DFT.

\FloatBarrier

\section{Acknowledgments}

We thank Leonardo dos Anjos Cunha, Min Lin, and Giovanni Vignale for valuable input. This work has been supported by the National Research Foundation, Singapore and A*STAR under its CQT Bridging Grant and its Quantum Engineering Programme (grant NRF2022-QEP2-02-P16 supports
J.H.H.). One of the authors (J.~C.) acknowledges the support of the Max-Planck-Institut f\"ur Physik komplexer Systeme, Dresden, Germany.

\appendix

\section{\label{AppendixPerturbationTheory}Perturbation theory for the correlation energy in the 1pEx-basis}

In this appendix we derive a perturbative approximation of the correlation contribution to the exact interaction energy. To that end, we calculate the contribution of the 2-cumulant---the difference between the exact and the HF two-body density matrix---to second-order perturbation theory along the lines of Ref.~\cite{Cioslowski2019}. This allows us, in the limit of weak interactions, to gauge how fast we approach the exact energies and participation numbers as we increase the number $L$ of single-particle states that determine the perturbative energies and participation numbers. Unfortunately, we find that the energies converge rather slowly with $L$, which suggests that the 1pEx-basis can be inefficient, compared with natural orbitals---at least as far as correlations beyond HF exchange are concerned, see also Ref.~\cite{Cioslowski2023}.

The expectation value of the many-particle Hamilton operator, cf.~Eq.~(\ref{eq:A8}), corresponding to some (not necessarily ground-state) wave function $\Psi$ reads
\begin{align}\label{9}
E = \langle\Psi | H_{\mathrm{mp}}|\Psi\rangle = \sum_{pq} {^{1} \Gamma}_{pq} \, h_{qp} + \sum_{pqrs} {^{2} \Gamma}_{pqrs} \, g_{rspq} \,,
\end{align} 
where $h_{pq}=\bok{p(1)}{H_{\mathrm{1p}}}{q(1)}$ and $g_{pqrs} = \bra{p(1)}\bok{q(2)}{V_{\mathrm{int}}(\vec R_1-\vec R_2)}{r(1)}\ket{s(2)}$, with $\ket{q(1)}$ referring to the spin-orbital $\phi_q(1)$ for a particle labeled `(1)'. That is, in what follows, the participation numbers are constrained by ${0\leq n_q\leq 1}$, and the spin-orbitals $\big\{\phi_q(\,)\big\}$ define the creation and annihilation operators $\big\{\phi_p^{+}\big\}$ and $\big\{\phi_p^{-}\big\}$ that enter the one-body reduced density matrix (`1-matrix') with elements
\begin{align}\label{a3}
{^{1} \Gamma_{pq}} = \langle \Psi | \phi_q^{+} \, \phi_p^{-} | \Psi \rangle \quad 
\end{align}     
and the 2-matrix with elements
\begin{align}\label{a5}
{^{2} \Gamma_{pqrs}} = \frac{1}{2} \, \langle \Psi | \phi_r^{+} \, \phi_s^{+} \, \phi_q^{-} \, \phi_p^{-}  | \Psi \rangle\,.
\end{align}
Their normalizations are
\begin{align}\label{13}
\sum_r \, {^{2} \Gamma_{prqr}} =  \frac{N-1}{2} \; {^{1} \Gamma_{pq}} \quad \mbox{and} \quad  \sum_p \, {^{1} \Gamma_{pp}} = N \,.
\end{align}
In conventional approaches to DFT, the total energy in Eq.~(\ref{9}) is the functional
\begin{align}\label{aa}
E[\rho(1)] = \min_{\{ {^{2} \Gamma_{pqrs}} \} \to \rho(1)} \;
\sum_{pqrs}  \,  \left( \frac{2 \, h_{rp} \, \delta_{qs}}{N-1} + g_{rspq} \right) \, {^{2} \Gamma_{pqrs}}  \, ,
\end{align}
where $\{ {^{2} \Gamma_{pqrs}} \} \to \rho(1)$, with the single-particle density $\rho(1)$, denotes both the constraint
\begin{align}\label{aa2}
\sum_{pq} {^{1} \Gamma}_{pq} \, \phi^{*}_q(1) \,  \phi_p(1) =\rho(1)  \quad 
\end{align}
and the $N$-representability of $\{ {^{2} \Gamma_{pqrs}} \}$. It is instructive to compare Eq.~(\ref{aa}) with its analog
\begin{align}\label{bb}
E\big(\{ ^{1} \Gamma_{pq} \}\big) =  \sum_{pq} {^{1} \Gamma}_{pq} \, h_{qp} + \min_{\{ {^{2} \Gamma_{pqrs}} \} \to \{ ^{1} \Gamma_{pq} \}} \;
\sum_{pqrs} {^{2} \Gamma}_{pqrs} \, g_{rspq}
\end{align}
in density matrix functional theory, where $\{ {^{2} \Gamma_{pqrs}} \} \to \{ ^{1} \Gamma_{pq} \}$ denotes both the constraints in Eq.~(\ref{13}) and the $N$-representability of $\{ {^{2} \Gamma_{pqrs}} \}$. There are two advantages of density matrix functional theory over conventional DFT. First, the obvious one, which is shared with 1pEx-DFT, namely the exclusion of the single-particle part from the constrained minimization. Second, the less obious one, which is only revealed upon the introduction of the 2-cumulant with the elements
\begin{align}\label{14}
^{2} \mathfrak{G}_{pqrs} = {^{2} \Gamma_{pqrs}} -\frac{1}{2} \left ( {^{1} \Gamma_{pr}} {^{1} \Gamma_{qs}} - {^{1} \Gamma_{ps}} {^{1} \Gamma_{qr}} \right ) \, ,
\end{align}
which obey ${^{2} \mathfrak{G}_{pqrs}={^{2}} \mathfrak{G}_{rspq}^{*} = - ^{2} \mathfrak{G}_{pqsr}  = {^{2} \mathfrak{G}_{qpsr}} = - ^{2}\mathfrak{G}_{qprs}}$ and
\begin{align}\label{16}
\sum_r \, {^{2} \mathfrak{G}_{prqr}} = \frac{1}{2} \left (\sum_k {^{1} \Gamma}_{pk}{^{1} \Gamma}_{kq} - {^{1} \Gamma}_{pq} \right ) \, .
\end{align}
Then, Eq.~(\ref{bb}) reads
\begin{align}\label{bbb}
E(\{ ^{1} \Gamma_{pq} \}) =  \sum_{pq} {^{1} \Gamma}_{pq} \, h_{qp} 
+  \frac{1}{2} \; \sum_{pqrs}\left ( {^{1} \Gamma_{pr}} {^{1} \Gamma_{qs}} - {^{1} \Gamma_{ps}} {^{1} \Gamma_{qr}} \right )  g_{rspq} + \min_{\{ {^{2} \mathfrak{G}_{pqrs}} \} \to \{ ^{1} \Gamma_{pq} \}} \; \sum_{pqrs} {^{2} \mathfrak{G}_{pqrs}} \, g_{rspq}  \, ,
\end{align}
where $\{ {^{2} \mathfrak{G}_{pqrs}} \} \to \{ ^{1} \Gamma_{pq} \}$ denotes both the constraint (\ref{16}) and the $N$-representability of $\{ {^{2} \mathfrak{G}_{pqrs}} \}$.  As a result, the minimization is now over a contribution to the total energy that vanishes for uncorrelated systems and is small in general.  

In 1pEx-DFT we choose the spin-orbitals $\big\{\phi_p(\,)\big\}$ specifically to diagonalize $H_{\mathrm{1p}}$. That is, each 1pEx mode introduced in Eq.~(\ref{eq:A3}) is effectively made up of two spin-orbitals $\phi_p(\,)$, and the total energy functional reads
\begin{align}\label{cc}
E(\{ n_{p} \}) =  \sum_{p} n_{p} \, h_{pp} + \min_{\{ {^{2} \Gamma_{pqrs}} \} \to \{ n_{p} \}} \;
\sum_{pqrs} {^{2} \Gamma}_{pqrs} \, g_{rspq}  \, ,
\end{align}
where $n_{p} = {^{1} \Gamma}_{pp}$, $\epsilon_{p} = h_{pp}$ (in the case of a spin-unpolarized system, ${\epsilon_{1}=\epsilon_{2}=E_1}$, ${\epsilon_{3}=\epsilon_{4}=E_2}$, and so forth, with the energies $E_a$ of Eq.~(\ref{eq:A3})), and $\{ {^{2} \Gamma_{pqrs}} \} \to \{ n_{p} \}$ denotes both the constraint ${\frac{2}{N-1} \;\sum_r \, {^{2} \Gamma_{prpr}} = n_{p}}$ for all $p$, according to Eq.~(\ref{13}), and the $N$-representability of $\{ {^{2} \Gamma_{pqrs}} \}$.

A perturbative treatment of the interaction energy in Eq.~(\ref{cc}) yields
\begin{align}\label{GammaExpansion}
{^{1} {\Gamma}_{pq}}={^{1} {\Gamma}_{pq}^{(0)}}+\lambda \, {^{1} {\Gamma}_{pq}^{(1)}}+\lambda^2 \, {^{1} {\Gamma}_{pq}^{(2)}}+...\,
\end{align}
with parameter $\lambda$, as done in Ref.~\cite{Cioslowski2019}, where
\begin{align}
{^{1} \Gamma_{pq}^{(0)}} &= \nu_p \, \delta_{pq} \, ,\label{27}\\
{^{1} \Gamma_{pq}^{(1)}} &=\frac{\nu_p-\nu_q} {\epsilon_p - \epsilon_q} \; G_{pq} \,  ,\label{28}
\end{align}
and
\begin{align}
{^{1} {\Gamma}_{pq}^{(2)}} &= \sum_i \, \frac{\eta_p \, \eta_q \, \nu_i-\nu_p \, \nu_q \,\eta_i}{(\epsilon_p - \epsilon_i) \, (\epsilon_q - \epsilon_i)} \; G_{pi} \; G_{iq} + \mathcal{\hat P}^{\dagger}_{pq} \sum_i \,  \frac{\eta_q \, \nu_p \, \nu_i-\nu_q \, \eta_p \,\eta_i} {\epsilon_q - \epsilon_i} \; \frac{G_{pi} \; G_{iq}} {\epsilon_p - \epsilon_q}\nn \\
&\quad+ \sum_{ij} \, \frac{\nu_p - \nu_q} {\epsilon_p - \epsilon_q} \; \frac{\nu_i - \nu_j}{\epsilon_i - \epsilon_j} \; G_{ij} \; \mathfrak{g}_{jpiq} 
+ \frac{1}{2} \mathcal{\hat P}^{\dagger}_{pq} \sum_{ijk} \, \frac{\eta_q \, \nu_p \, \nu_i \, \nu_j \, \eta_k - \nu_q \, \eta_p \, \eta_i \, \eta_j \, \nu_k}{\epsilon_q+ \epsilon_k - \epsilon_i - \epsilon_j} \; \frac{\mathfrak{g}_{kpij} \; \mathfrak{g}_{ijkq}} {\epsilon_p - \epsilon_q}\nn \\
&\quad+ \, 2 \, (\eta_p \, \eta_q - \nu_p \, \nu_q) \sum_{ijk} \, {^{2} {\mathfrak{G}}_{ipjk}^{(1)}} \, {^{2} {\mathfrak{G}}_{jkiq}^{(1)}}  \,.\label{aa3a}
\end{align}
Here, $\nu_p$ and ${\eta_p=1-\nu_p}$ are the indicator functions for the core ($\{\nu_p\}$) and virtual ($\{\eta_p\}$) subsets of the spin-orbitals $\{\phi_p(1)\}$, i.e., ${\nu_p=1}$ if $n_p$ is close to one and ${\nu_p=0}$ otherwise. For the purpose of the present work, we set ${\nu_p=1}$ for ${p=1,...,N}$ in the case of $N$ particles. Fractions with vanishing numerators in Eqs.~(\ref{28}) and (\ref{aa3a}) are assumed to be zero, even if ${\epsilon_p=\epsilon_q}$. We define ${\mathfrak{g}_{pqrs} = g_{pqrs}-g_{pqsr}}$ and ${G_{pq} = \sum_i \, \nu_i \, \mathfrak{g}_{piqi}}$, and the operator $\mathcal{P}^{\dagger}_{pq}$ acts on two-index quantities according to the prescription ${\mathcal{P}^{\dagger}_{pq} \, A_{pq} = A_{pq}+A_{qp}^{*}}$. 

The 2-cumulant has no zeroth-order term and we shall approximate it by its first-order term
\begin{align}\label{29}
\lambda\; ^{2} \mathfrak{G}_{pqrs}^{(1)} = \frac{1}{2}  \; \frac{\nu_p \, \nu_q \, \eta_r \, \eta_s - \eta_p \, \eta_q \, \nu_r \, \nu_s}{\epsilon_p+ \epsilon_q - \epsilon_r - \epsilon_s} \; \mathfrak{g}_{pqrs} \, ,
\end{align}
see Ref.~\cite{Cioslowski2019}. With Eqs.~(\ref{14}), (\ref{GammaExpansion}), and (\ref{29}) the first two terms in the expansion 
\begin{align}
V_{\mathrm{ee}}(\{ n_{p} \}) =  \min_{\{ {^{2} \Gamma_{pqrs}} \} \to \{ n_{p} \}} \; \sum_{pqrs} {^{2} \Gamma}_{pqrs} \, g_{rspq} = \lambda \, V^{(1)}_{\mathrm{ee}}(\{ n_{p} \})+\lambda^2 \, V_{\mathrm{ee}}^{(2)}(\{ n_{p} \})+...
\end{align}
are readily available:
\begin{align}\label{zz1}
V_{\mathrm{ee}}^{(1)}(\{ n_{p} \}) &= \sum_{pqrs} {^{2} \Gamma}^{(0)}_{pqrs} \, g_{rspq} = \frac{1}{2} \; \sum_{pqrs} \left ( {^{1} \Gamma^{(0)}_{pr}} \, {^{1} \Gamma^{(0)}_{qs}} - {^{1} \Gamma^{(0)}_{ps}} \, {^{1} \Gamma^{(0)}_{qr}} \right ) \, g_{rspq} = \frac{1}{2} \; \sum_{pq} \, \nu_p \, \nu_q \, \mathfrak{g}_{pqpq}\\
\intertext{and}
V_{\mathrm{ee}}^{(2)}(\{ n_{p} \}) &= \sum_{pqrs} {^{2} \Gamma}^{(1)}_{pqrs} \, g_{rspq} \nn\\
&= \frac{1}{2} \; \sum_{pqrs} \left ( {^{1} \Gamma^{(1)}_{pr}} \, {^{1} \Gamma^{(0)}_{qs}} - {^{1} \Gamma^{(1)}_{ps}} \, {^{1} \Gamma^{(0)}_{qr}} + {^{1} \Gamma^{(0)}_{pr}} \, {^{1} \Gamma^{(1)}_{qs}} - {^{1} \Gamma^{(0)}_{ps}} \, {^{1} \Gamma^{(1)}_{qr}}\right ) \, g_{rspq} + \, \sum_{pqrs}  {^{2} \mathfrak{G}_{pqrs}^{(1)}} \, g_{rspq}  \nn \\
&= \, \sum_{pq}  \frac{\nu_p-\nu_q} {\epsilon_p - \epsilon_q} \; |G_{pq}|^2 +  \frac{1}{2} \, \sum_{pqrs} \frac{\nu_p \, \nu_q \, \eta_r \, \eta_s - \eta_p \, \eta_q \, \nu_r \, \nu_s}{\epsilon_p+ \epsilon_q - \epsilon_r - \epsilon_s} \; |\mathfrak{g}_{pqrs}|^2  \, ,
\end{align}
such that ($\lambda$ is now dropped)
\begin{align}\label{gg}
V_{\mathrm{ee}}(\{ n_{p} \}) = \frac{1}{2} \; \sum_{pq} \, \nu_p \, \nu_q \, \mathfrak{g}_{pqpq} + \, \sum_{pq}  \frac{\nu_p-\nu_q} {\epsilon_p - \epsilon_q} \; |G_{pq}|^2 +  \frac{1}{2} \, \sum_{pqrs} \frac{\nu_p \, \nu_q \, \eta_r \, \eta_s - \eta_p \, \eta_q \, \nu_r \, \nu_s}{\epsilon_p+ \epsilon_q - \epsilon_r - \epsilon_s} \; |\mathfrak{g}_{pqrs}|^2 
\end{align}
is correct up to second order in the pair interaction strength. Accordingly, we determine the participation numbers by setting ${p=q}$ in Eqs.~(\ref{27})--(\ref{aa3a}): 
\begin{align}\label{aa3c}
n_p = {^{1} \Gamma}_{pp} \approx \nu_p + \sum_i \, \frac{\eta_p  \, \nu_i-\nu_p  \,\eta_i} {(\epsilon_p - \epsilon_i)^2} \; |G_{pi}|^2 + \, \frac{1}{2} \; (\eta_p - \nu_p) \, \sum_{ijk} \frac{\eta_p \, \eta_i \, \nu_j \, \nu_k+ \nu_p \, \nu_i \, \eta_j \, \eta_k}{(\epsilon_i+ \epsilon_p - \epsilon_j - \epsilon_k)^2} \; |\mathfrak{g}_{ipjk}|^2\,.
\end{align}
\begin{figure}[ht]
\includegraphics[width=\linewidth]{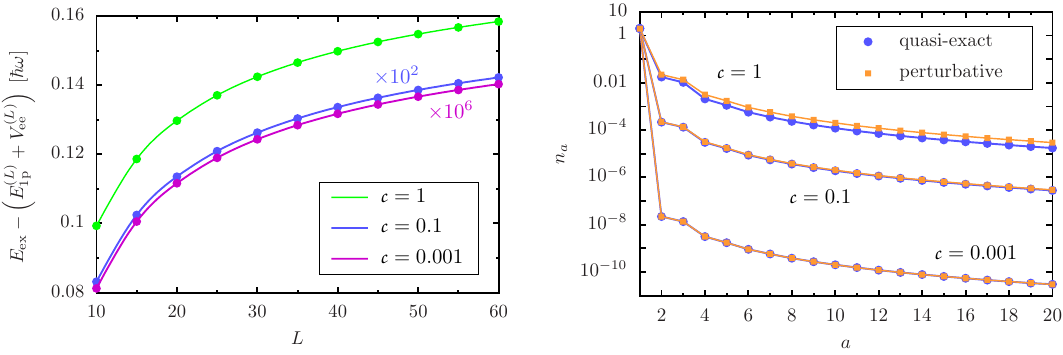}
\caption{\label{FigurePerturbationTheory}Energies and participation numbers for ${N=2}$ contact-interacting unpolarized spin-$\frac12$ particles at three different interaction strengths ${\mathpzc{c}}$. The connecting lines guide the eye. Left: we compare the quasi-exact energy $E_{\mathrm{ex}}$ (where `quasi-exact' refers to a numerically approximate evaluation of the exact expression that is accurate at the level of machine precision, here achieved for ${L=60}$) with the total energy ${E_{\mathrm{1p}}+V_{\mathrm{ee}}}$ (as a function of $L$) to second-order perturbation theory, viz., Eq.~(\ref{gg}), with $2L$ spin-orbitals taken into account. For illustrative purposes we rescale the ordinate for ${\mathpzc{c}=0.1}$ (${\mathpzc{c}=0.001}$) by a factor of $10^2$ ($10^6$). Right: the participation numbers $\{n_a\}$ of (doubly occupied) 1pEx modes labeled ${a=1,2,...,L}$ according to Eq.~(\ref{eq:A3}). We compare the quasi-exact participation numbers with the participation numbers derived from Eq.~(\ref{aa3c}), where the summation indices span~$\{1,2,...,2L\}$.}
\end{figure}
Figure~\ref{FigurePerturbationTheory} illustrates for ${N=2}$ contact-interacting unpolarized spin-$\frac12$ particles that the total energy in second-order perturbation theory converges rather slowly as the number $L$ of single particle orbitals increases (Fig.~\ref{FigurePerturbationTheory}, left panel). As expected, the perturbative treatment becomes very accurate as the contact interaction strength $\mathpzc{c}$ decreases from ${\mathpzc{c}=1}$ to ${\mathpzc{c}=0.001}$. But the convergence behavior for ${\mathpzc{c}=0.001}$ still matches that for ${\mathpzc{c}=1}$. Similarly, we find a slow decay of the 1pEx participation numbers $\{n_a\}$ toward zero as $a$ increases (Fig.~\ref{FigurePerturbationTheory}, right panel). If this example is prototypical, then a high energy cutoff in the 1pEx-basis is required for very precise calculations that include correlations beyond HF exchange.

\FloatBarrier

\section{\label{AppendixApproximateRDMs}Approximate density matrices}

In this appendix we (i) obtain the seed matrices for the 1pEx-DFT program in Eq.~(\ref{eq:D5}) from the matrix mixer algorithm introduced in Ref.~\cite{ParaniakPhdThesis2022} and from a TF-inspired Wigner function, respectively, (ii) derive the Hartree--Fock density matrix in the 1pEx-basis for ${N=2}$ particles, (iii) express the spatial and momental densities through the converged seed matrix and phases, (iv) write the exact HF interaction energy for the contact-interaction as a density functional, and (v) discuss transformations of the seed matrix beyond the phase transformations in Eq.~(\ref{eq:C6}).\\

\textbf{Seed matrix from a matrix mixer algorithm.} To generate a valid density matrix $\varrho$, which obeys
\begin{align}
    \label{eq:rho-cons-1}
    \varrho_{aa} &= n_a 
    \intertext{and}
    \label{eq:rho-cons-2}
    \varrho^2 &= 2 \varrho
\end{align}
for a prescribed vector $\dens{n}$ of participation numbers, we start from a matrix $\varrho_0$ with diagonal entries $(2,2,\dots,2,0,0,\dots,0)$ that add up to $N$. Note that $\varrho_0$ is itself a proper density matrix and obeys the constraint of Eq.~(\ref{eq:rho-cons-2}). We aim at an iterative transformation of $\varrho_0$ by unitary transformations (leaving both the trace and the spectrum unchanged), such that the diagonal of the final matrix is $\dens{n}$. This is accomplished by a sequence of unitary transformations
\begin{equation}
    M = \begin{pmatrix}
        \cos\theta & \sin\theta \\ 
        \sin\theta & -\cos\theta \\
    \end{pmatrix}\,,
\end{equation}
which transform diagonal $2\times2$ matrices into
\begin{equation}
    \label{eq:mixing-operation}
    M(\eta) \,
\begin{pmatrix}
    a & 0 \\ 
    0 & b
\end{pmatrix}\,M(\eta)^\dagger =
\begin{pmatrix}
\eta a + (1-\eta)b & \sqrt{\eta (1-\eta)} (a-b) \\
\sqrt{\eta (1-\eta)} (a-b) & (1-\eta)a + \eta b
\end{pmatrix}.
\end{equation}
Here, the parameter ${\eta = \cos^2\theta}$ can be adjusted for the first diagonal element of the matrix in Eq.~(\ref{eq:mixing-operation}) to equal any prescribed value between $a$ and $b$. In particular, since the participation numbers are all bounded between zero and two, we can apply a sequence of suitable transformations $M$ on the initial matrix $\varrho_0$, such that the final density matrix $\varrho^{\mathrm{it}}$ obeys the constraint in Eq.~(\ref{eq:rho-cons-1}).

The matrix mixer algorithm has been proven to work in all cases \cite{ParaniakPhdThesis2022}. The proof consists of inductively showing that, for a given $n_a$, there is always a $2\times 2$ diagonal sub-matrix, associated with indices $(i,j)$, such that ${\varrho_{ii} \ge n_a \ge \varrho_{jj}}$, hence permitting the application of the matrix mixing operation $M$ of Eq.~(\ref{eq:mixing-operation}). As an illustration, we consider ${N=4}$ particles on ${L=4}$ levels and target ${\dens{n} = \left(\frac{6}{4}, \frac{5}{4}, \frac{3}{4}, \frac{2}{4}\right)}$. The sequence
\begin{equation}
    \label{eq:mm-example-l4}
        \varrho_0 = \begin{pmatrix}
        2 & 0 & 0 & 0 \\
        0 & 2 & 0 & 0 \\
        0 & 0 & 0 & 0 \\
        0 & 0 & 0 & 0 \\
    \end{pmatrix}
        \overset{\eta_1 = 3/4}{\underset{(1,3)}{\longrightarrow}}\; \varrho_1 = \begin{pmatrix}
            \frac{3}{2} &  & * &  \\
         & 2 &  &  \\
         * &  & \frac{1}{2} &  \\
         &  &  &  0 \\
    \end{pmatrix}
        \overset{\eta_2 = 1/2}{\underset{(2,3)}{\longrightarrow}}\; \varrho_2 = \begin{pmatrix}
        \frac{3}{2} & * & * &  \\
            * & \frac{5}{4}  & *  &  \\
            * & * & \frac{5}{4} &  \\
         &  &  & 0 \\
        \end{pmatrix}
        \overset{\eta_3 = 3/5}{\underset{(3,4)}{\longrightarrow}}\; \varrho_3 = \begin{pmatrix}
        \frac{3}{2} & * & * & * \\
            * & \frac{5}{4} & * & * \\
        * & * & \frac{3}{4} & * \\
        * & * & * & \frac{1}{2} \\
        \end{pmatrix}
\end{equation}
of matrix mixing operations $M(\eta_i)$, each operating in the $2\times 2$ sector of indices $(a,b)$ as indicated, yields one target entry on the diagonal in each step $i$. Off-diagonal elements, here summarily denoted by an asterisk ($*$), are thereby introduced and modified. The final density matrix $\varrho_3$ then satisfies both conditions of Eqs.~(\ref{eq:rho-cons-1}) and (\ref{eq:rho-cons-2}). The exploration of alternative mixer algorithms is left for future study.

In the simplest case of ${N=2}$ particles on ${L=2}$ levels, i.e., ${n_a + n_b = 2}$ (without loss of generality, ${n_a \ge n_b}$), we begin with ${\varrho_0 = \begin{pmatrix} 2 & 0 \\ 0 & 0 \\ \end{pmatrix}}$ and apply Eq.~(\ref{eq:mixing-operation}) with weight $\eta = \frac{n_a}{2}$ to obtain
\begin{equation}
    \varrho^{\mathrm{it}} = \begin{pmatrix}
        n_a & \sqrt{n_a n_b} \\
        \sqrt{n_a n_b} & n_b 
    \end{pmatrix}.
\end{equation}
That is, for ${N=2}$ particles, the matrix mixer algorithm reproduces the Hartree--Fock density matrix in the 1pEx-basis: the spin-orbital density matrix $\gamma_{\mathrm{HF}}^{N=2}(r;r')$ of the Hartree--Fock spin-singlet ground-state wave function
\begin{align}
\psi_{\mathrm{HF}}^{N=2}\big(r_1=(\vec r_1,\sigma_1),r_2=(\vec r_2,\sigma_2)\big)=\frac{1}{\sqrt{2}}\phi(\vec r_1)\,\phi(\vec r_2)\,\bra{\sigma_1\sigma_2}\,\big(\ket{\uparrow\downarrow}-\ket{\downarrow\uparrow}\big)
\end{align}
for two particles in the normalized spatial orbital $\phi(\,)$ is
\begin{align}\label{gammaHF}
\gamma_{\mathrm{HF}}^{N=2}(r;r')=2\int\d r_2\,\psi_{\mathrm{HF}}^{N=2}(r,r_2)\,\psi_{\mathrm{HF}}^{N=2}(r',r_2)^*=\phi(\vec r)\,\phi(\vec r')^*\,\big(\bk{\sigma}{\uparrow}\bk{\uparrow}{\sigma'}+\bk{\sigma}{\downarrow}\bk{\downarrow}{\sigma'}\big)\,.
\end{align}
Upon integrating out the spin degrees of freedom, we obtain the orbital density matrix
\begin{align}
\gamma_{\mathrm{HF}}^{N=2}(\vec r;\vec r')=\sum_{\sigma,\sigma'\in\{\uparrow,\downarrow\}}\gamma_{\mathrm{HF}}^{N=2}(r;r')=\bok{\vec r}{\rho_{\mathrm{HF}}^{N=2}}{\vec r'},
\end{align}
with ${\rho_{\mathrm{HF}}^{N=2}=2\ket{\phi}\bra{\phi}}$, such that $\varrho^{\mathrm{HF},N=2}_{ab}=\bok{a}{\rho_{\mathrm{HF}}^{N=2}}{b}=2\,c_a\,c_b^*$, with coefficients ${c_a=|c_a|\,\mathrm{e}^{\I\phi_a}}$ of the expansion ${\phi(\vec r)=\sum_ac_a\psi_a(\vec r)}$ in the orthonormal 1pEx-basis. Since ${\varrho^{\mathrm{HF},N=2}_{aa}=2|c_a|^2=n_a}$, we get
\begin{align}                                                                                                                                                                                                                                                                                                                                                                       
\varrho^{\mathrm{HF},N=2}_{ab}=\sqrt{n_a\,n_b}\,\mathrm{e}^{\I\,\big(\phi_a-\phi_b\big)}\equiv\varrho^{\mathrm{it},N=2}_{ab}=\Exp{\I\phi_{a}}\varrho^{(0),\mathrm{it},N=2}_{ab}\Exp{-\I\phi_{b}},                                                                                                                                                                                                                                                                                                                                                           \end{align}
with optimal phases $\{\hat{\phi}_a\}$ to be found, see Eq.~(\ref{eq:D5}).

\begin{figure}[ht]
\begin{center}
\includegraphics[width=0.65\linewidth]{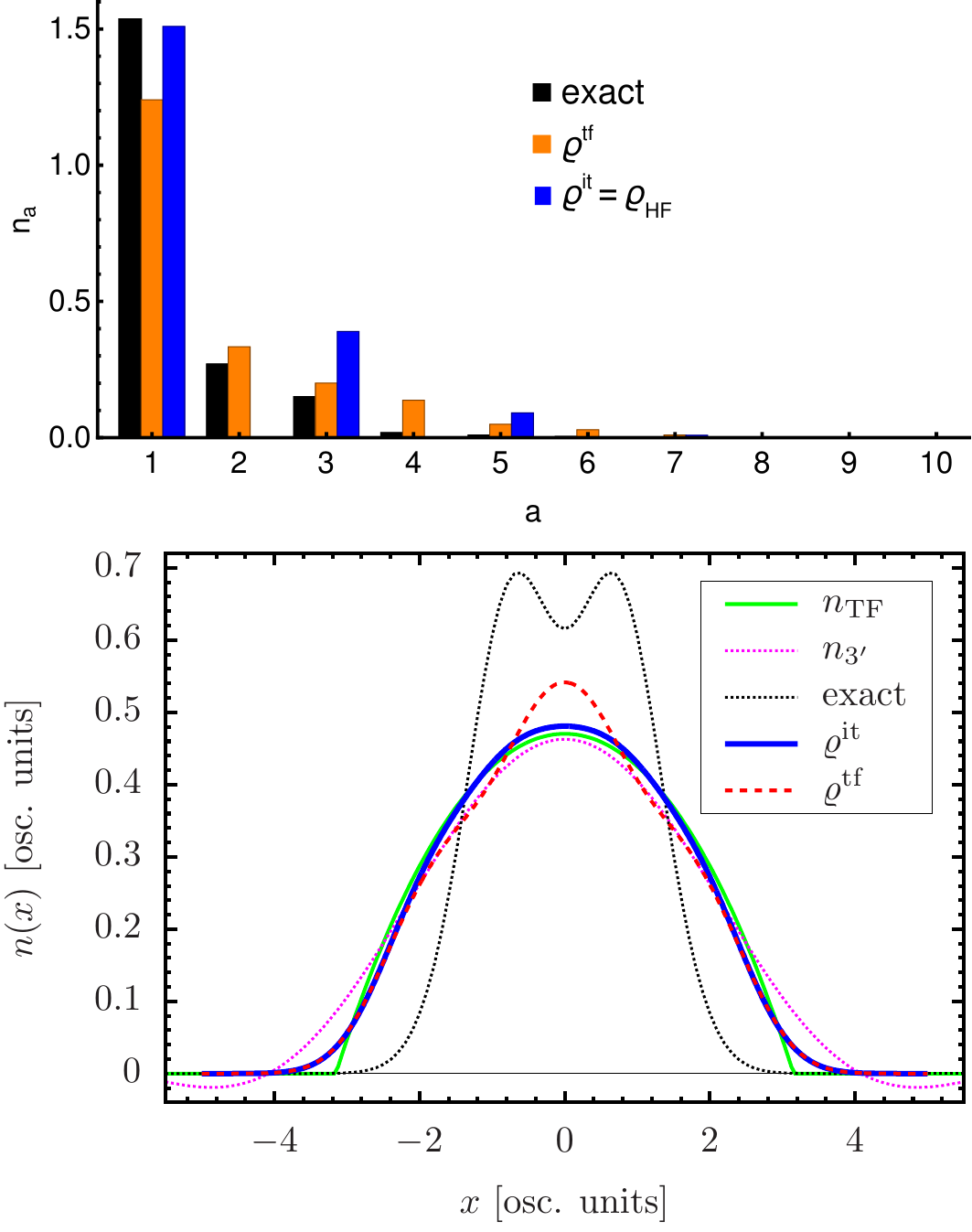}
\caption{\label{1Dcontact_OccNum_2} Participation numbers $n_a$ for single-particle levels ${a=1,\dots,L=10}$ (top) and spatial densities (bottom) for ${N=2}$ spin-$\half$ fermions in a 1D harmonic trap at contact-interaction strength ${\mathpzc{c}=20}$, analogous to Fig.~\ref{1Dcontact_OccNum_10}. We obtain the participation numbers from 1pEx-DFT using the HF density matrix, which coincides with $\varrho^{\mathrm{it}}$ for ${N=2}$. For this system, HF overestimates the exact ground-state energy by a factor of three. Hence, the quantitative differences between the (Dirac-approximated) participation numbers and their exact counterparts obtained from Refs.~\cite{Busch+3:98, Viana-Gomes+1:11} are not surprising. The same holds for the spatial 1pEx-densities labeled `$\varrho^{\mathrm{it/tf}}$', which align more with the DPFT densities $n_{\mathrm{TF}}$ and $n_{3'}$ than with the exact density. As any approximate density, also $n_{3'}$ does not have all the properties of the exact density. The oscillations of $n_{3'}$ into negative numbers occur in the classically forbidden region and become less pronounced for larger $N$, see Fig.~\ref{1Dcontact_OccNum_10} and Ref.~\cite{Trappe2023DFMPS}. For applications that strictly rely on positive densities everywhere, positive $n_{3'}$ can be enforced in an ad hoc fashion at the level of the self-consistent calculation that yields $n_{3'}$, see Ref.~\cite{Trappe2021b}. A similar measure is standard in the TF approach, where the spatial density is manually made to vanish in the classically forbidden region. We report exact results for completeness, but our actual objective is to validate our 1pEx-DFT implementation by comparing with HF results.}
\end{center}
\end{figure}

In Fig.~\ref{1Dcontact_OccNum_2} we compare the optimized participation numbers for this case with the exact results and display the spatial densities calculated from the converged density matrix: if the 1pEx-basis states can be chosen real, as is the case for the harmonic oscillator eigenstates, the spatial (momental) densities are 
\begin{align}\label{spatialmomentaldensity}
n(\cdot)=\sum_{a,b}\cos\big(\hat{\phi}_a-\hat{\phi}_b\big)\,\psi_a(\cdot)\,\hat{\varrho}_{ab}^{(0)}\,\psi_b(\cdot)\,, 
\end{align}
where $(\cdot)$ stands for position $\vec r$ (momentum $\vec p$). We can also use Eq.~(\ref{spatialmomentaldensity}) in the case of the hydrogenic states after separating from $\{\psi_a\}$ the $\phi$-dependence (with corresponding quantum number $m_a$) of the spherical harmonics in Eq.~(\ref{HydrogenicWaveFunction}), such that the remainder of $\{\psi_a\}$ is real, and adding $\phi\,(m_a-m_b)$ to the argument of the cosine in Eq.~(\ref{spatialmomentaldensity}), with $\phi$ the azimuthal angle of $\vec r$ (or $\vec p$).

Using the density matrix in Eq.~(\ref{gammaHF}) and the pair potential $V_{\mathrm{int}}$ in Eq.~(\ref{VintContact}), we reduce the exact HF interaction energy
\begin{align}
E_{\mathrm{int}}^{\mathrm{HF}}[\gamma]=\frac12\int(\d\vec r_1)(\d\vec r_2)\,V_{\mathrm{int}}(\vec r_1-\vec r_2)\left(n(\vec r_1)n(\vec r_2)-\sum_{\sigma_1,\sigma_2}\gamma_{\mathrm{HF}}^{N=2}(r_1;r_2)\,\gamma_{\mathrm{HF}}^{N=2}(r_2;r_1)\right)
\end{align}
to
\begin{align}\label{EintHFN2}
E_{\text{int,1D-contact}}^{\mathrm{HF}}[n]=\frac{\mathpzc{c}}{2}\int\d x\,\left(\big(n(x)\big)^2-\frac12\big(n(x)\big)^2\right)=\frac{\mathpzc{c}}{4}\int\d x\,\big(n(x)\big)^2\,,
\end{align}
which is actually the exact expression of the HF interaction energy for any even $N$ (and generalizes to contact-interaction in 2D and 3D). We use Eq.~(\ref{EintHFN2}) as the interaction functional for producing the DPFT energies in Table~\ref{1DcontactTable} and the DPFT densities in Figs.~\ref{1Dcontact_OccNum_10} and \ref{1Dcontact_OccNum_2}.\\

\textbf{Seed matrix from a Thomas--Fermi inspired density matrix.}  With the ingredients listed in Table~\ref{TableRotor}, we produce the density matrix $\varrho^{\mathrm{tf}}$ in Eq.~(\ref{rhotf}) in analogy to the TF-approximated spatial density matrix through an ansatz for an approximate Wigner function $[F]_W$ of an operator $F$ in the rotor phase space. The basic observables of the rotor are (i) the unitary ${\Varepsilon=\Exp{\I\Phi}}$ for the azimuth with bras ${\bra{\varphi}=\bra{\varphi+2\pi}}$ and $\Varepsilon\ket{\varphi}=\Exp{\I\varphi}\ket{\varphi}$, and (ii) the hermitian $\mathcal A$ for the angular momentum with kets $\ket{a}$. The inversion operator is ${I=\sum_{a=-\infty}^\infty\ket{-a}\bra{a}=\int_{(2\pi)}\frac{\d\varphi}{2\pi}\ket{-\varphi}\bra{\varphi}}$, $\eta(\,)$ is the step function, and ${\nu=\sqrt{\frac{2m}{\hbar^2}\left[\mu-V\left(\frac{x+x'}{2}\right)\right]_+}}$, with chemical potential $\mu$, potential energy $V(\,)$, and ${[z]_+=z\,\eta(z)}$. We express $F(\mathcal A,\Varepsilon)$ with the help of the Fourier representation of $\eta(\,)$, where the contour integration crosses the imaginary axis in the lower half-plane, and define $f(\mathcal A)\ket{a}=f(a)\ket{a}=\cot\left(\frac{\pi}{2}n_a\right)\ket{a}$. The operators $P$ and $X$ that appear in $W(x,p;X,P)$ as powers of $(P;X)$ are ordered such that all $P$ stand to the left of $X$.

\FloatBarrier

\begin{center}
\begin{table}[ht]
\setlength\extrarowheight{0.9em}
\setlength\tabcolsep{0.55em}
\begin{center}
\begin{tabular}{r|l}
 & \parbox{18.01em}{classical phase space ${\left\{x\in\mathbb R,p\in\mathbb R\right\}}$\\[0em] rotor phase space ${\left\{a\in\mathbb Z,\varphi\in[-\pi,\pi)\right\}}$}\\[0.7em]
\hline\hline
operator basis of Wigner type & \parbox{18.01em}{$W(x,p;X,P)=2\,\Exp{\frac{2\I}{\hbar}(P-p);(X-x)}$\\ $W(a,\varphi;\mathcal A,\Varepsilon)=\Varepsilon^{a}\,\Exp{-\I\varphi\mathcal A}\,I\,(1+\Varepsilon)\,\Exp{\I\varphi\mathcal A}\,\Varepsilon^{-a}$}\\[1.5em]
operator representation & \parbox{18.01em}{$F(X,P)=\int\frac{\d X\d P}{2\pi\hbar}[F]_W(x,p)\,W(x,p;X,P)$\\[0.3em] $F(a,\varphi)=\sum_{a}\int_{(2\pi)}\frac{\d\varphi}{2\pi}[F]_W(a,\varphi)\,W(a,\varphi;\mathcal A,\Varepsilon)$}\\[1.1em]
ansatz & \parbox{18.01em}{${[F]_W(x,p)=g\,\eta\left(\mu-\frac{1}{2m}p^2-V(x)\right)}$\\[0.1em] ${[F]_W(a,\varphi)=g\,\eta\big(\cot\varphi-f(a)\big)}$}\\[1.5em]
operator corresponding to $[F]_W$ & \parbox{18.01em}{${F(X,P)=g\,\eta\left(\mu-\frac{1}{2m}P^2-V(X)\right)}$\\[0.3em] ${F(\mathcal A,\Varepsilon)=g\,\Int\frac{\d t}{2\pi\I t}\,\Exp{\I t\,[\cot\Phi-f(\mathcal A)]}}$}\\[1.3em]
approximate density matrix & \parbox{18.01em}{${\varrho_{\mathrm{TF}}(x;x')=\bok{r}{F}{r'}=\frac{g\sin\left[2(x-x')\,\nu\right]}{2\pi(x-x')}}$\\[0.3em] ${\varrho^{\mathrm{tf}}_{ab}=\bok{a}{F}{b}=\frac{g\sin\left[(a-b)\,\sigma\right]}{\pi(a-b)}=\mbox{Eq.~(\ref{rhotf})}}$}\\
\end{tabular}
\end{center}
\caption{\label{TableRotor} Key ingredients for the derivation of the density matrix $\varrho^{\mathrm{tf}}$ in rotor phase space---in analogy to the TF-approximated density matrix $\varrho_{\mathrm{TF}}$ in classical phase space, see also Refs.~\cite{Trappe2016,Trappe2017,Chau2018}.}
\end{table}
\end{center}

As an example, we give $\varrho^{\mathrm{it}}$ and $\varrho^{\mathrm{tf}}$ for ${\dens{n}=(2,1.5,0.3,0.2)}$:
\begin{align}
\varrho^{\mathrm{it}}
\approx\left(\begin{array}{cccc}
2 & 0 & 0 & 0 \\
0 & 1.5 & 0.67 & 0.55 \\
0 & 0.67 & 0.3 & 0.24 \\
0 & 0.55 & 0.24 & 0.2 
\end{array}\right)
\quad,\qquad
\varrho^{\mathrm{tf}}
\approx\left(\begin{array}{cccc}
2 & 0 & 0 & 0 \\
0 & 1.5 & 0.57 & 0.32 \\
0 & 0.57 & 0.3 & 0.23 \\
0 & 0.32 & 0.23 & 0.2 
\end{array}\right)\ .
\end{align}

\FloatBarrier

\textbf{Alternative transformations of the seed matrix.} We consider a $2\times2$ sector at the diagonal
of $\varrho$ (in general, the $2\times2$ sector associated with an index pair $(a,b)$), of the form
\begin{align}
  \label{eq:X1}
  \column[cccc]{ \ddots & \vdots & \vdots & \\
    \cdots & x & \gamma & \cdots\\
    \cdots & \gamma^* & y & \cdots\\
           & \vdots & \vdots & \ddots }\,,
\end{align}
where the unitary transformation afforded by
\begin{align}
  \label{eq:X2}
  U=\left[(x-y)^2+\left(\gamma^*\Exp{\I\varphi_{ab}}+\gamma\Exp{-\I\varphi_{ab}}\right)^2\right]^{-\half}
    \column[ccc]{(x-y)\,\Exp{\I\xi_a} && \big(\gamma+\gamma^*\Exp{2\I\varphi_{ab}}\big)\,\Exp{\I\xi_b} \\
      \big(\gamma^*+\gamma\Exp{-2\I\varphi_{ab}}\big)\,\Exp{\I\xi_a} && (y-x)\,\Exp{\I\xi_b}}
\end{align}
yields
\begin{align}
  \label{eq:X3}
  U^\dagger\,\column[cc]{x & \gamma \\ \gamma^* & y}\,U=
  \column[cc]{x & \gamma^*\Exp{\I(2\varphi_{ab}-\xi_a+\xi_b)} \\ \gamma\Exp{-\I(2\varphi_{ab}-\xi_a+\xi_b)} & y}\,,
\end{align}
so that the diagonal entries are unchanged, while the off-diagonal entries are affected---in fact, Eq.~(\ref{eq:X2}) parameterizes all $2\times2$ unitary transformations with this property. Our numerical analyses show, however, that the inclusion of the $3L(L-1)/2$ parameters $\{\varphi_{ab},\xi_a,\xi_b\}$, three for each index pair $(a,b)$, does not yield lower energies in the global minimization of Eq.~(\ref{eq:A9}) compared with optimizing over the $L$ phases of Eq.~(\ref{eq:C6}) only. Although every unitary transformation that leaves the diagonal elements of a matrix unchanged can be decomposed into a sequence of $L(L-1)/2$ transformations in $2\times2$ sectors, we have so far only covered those cases in which each of these transformations in $2\times2$ sectors individually preserve the diagonal elements. We leave the exploration of more general unitary transformations for future study.

\section{\label{AppendixEA}Evolutionary algorithms}

\subsection{Particle swarm optimization}

Particle swarm optimization (PSO) is an evolutionary algorithm that draws inspiration from how an `intelligent' swarm, such as a flock of birds, moves toward beneficial conditions \cite{Kennedy1995,Bonyadi2017,Tang2021}. Aiming at the global optimizer ${\hat{\vec x}=\mathrm{arg}\,\underset{\vec x\in X}{\mathrm{min}}f(\vec x)}$ of the objective function $f$, each swarm particle ${s\in \{1,\dots,S\}}$ updates its coordinate vector ${\vec x_s=\left(x_{s,1},\dots,x_{s,d},\dots,x_{s,D}\right)\in X}$ in an iterative random walk through the $D$-dimensional search space $X$. The core principle of PSO is the stochastic guidance of this random walk by (i) the so far encountered personal best position $\vec p_s$ of $s$ and (ii) the global personal best position $\vec p_{g}$---or, alternatively and more generally, the personal best position $\vec p_{g_s}$ from a randomly selected group $G_s$ of particles that are intermittently linked to $s$.

\begin{figure}[htb!]
\includegraphics[width=0.95\linewidth]{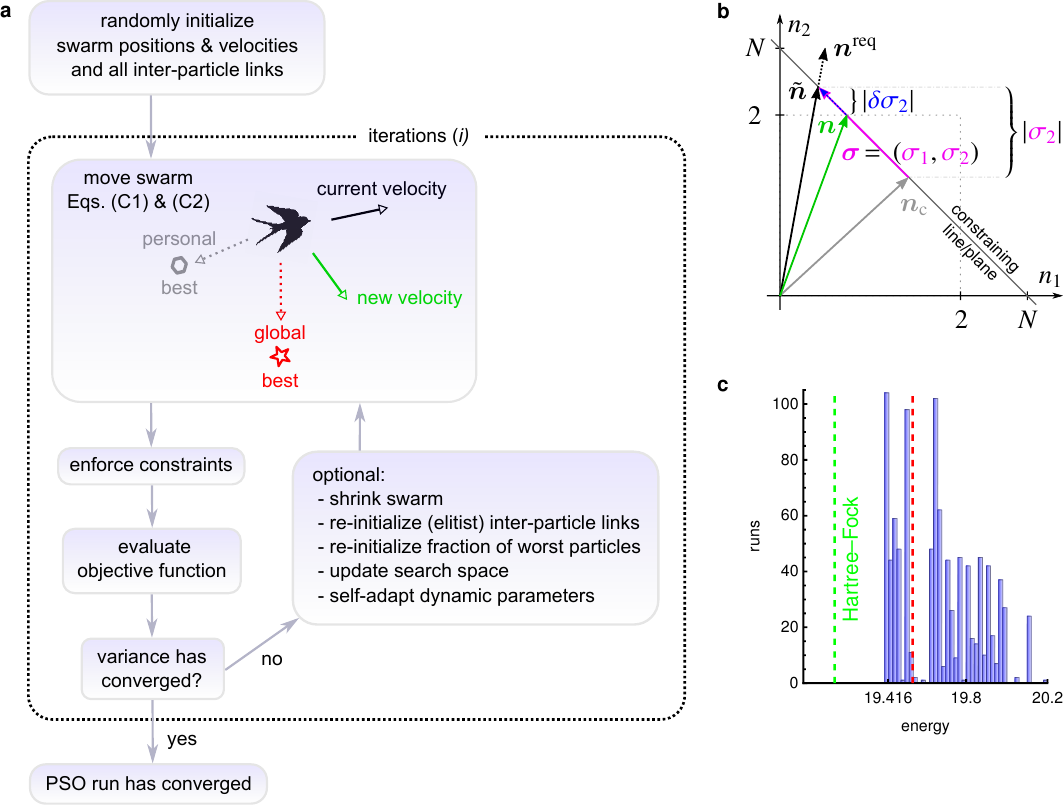}
\caption{\label{FigurePSO} \textbf{a}, Schematics of our implementation of PSO. \textbf{b}, An illustration of how to enforce the particle number constraint ${\sum_{i=1}^Ln_i=N}$ in the case of ${L=2}$ levels by replacing an improper $\dens{n}^{\mathrm{req}}$, the participation numbers for which the optimizer requests a function value, with ${\dens{n}=\dens{n}_{\mathrm c}+\big(\tilde{\dens{n}}-\dens{n}_{\mathrm c}\big)\big||\sigma_2|-|\delta \sigma_2|\big|/|\sigma_2|}$, which obeys all constraints and is a proper vector of participation numbers close to $\dens{n}^{\mathrm{req}}$. \textbf{c}, Histogram of 1000 PSO runs (totaling ${\sim10^8}$ energy evaluations) for ${N=4}$ harmonically confined spin-1/2 fermions with contact interaction strength ${\mathpzc{c}=20}$ on 20 single-particle levels (hence, the dimension of the search space is ${D=40}$), cf.~Sec.~\ref{Sec1DContact} and Table~\ref{1DcontactTable}. The lowest (highest) converged energy found by PSO is $19.416$ ($21.564$). About $36\%$ of the runs yield energies below the red dashed line, which marks two percent excess over the HF energy of 19.154 (green dashed line).}
\end{figure}

Figure~\ref{FigurePSO}a illustrates the work flow of our PSO implementation. We start each PSO run by uniformly drawing $S$ random particle `positions' $\vec x_s^{(0)}$ and particle `velocities' $\vec v_s^{(0)}$ from $X$. Then, in the $i$th iteration of the run, each position coordinate $x_{s,d}^{(i-1)}$ receives the update \begin{align}\label{PSOxUpdate}
x_{s,d}^{(i-1)}\longrightarrow x_{s,d}^{(i)}=x_{s,d}^{(i-1)}+v_{s,d}^{(i)}
\end{align}
according to the velocity update
\begin{align}\label{PSOvUpdate}
v_{s,d}^{(i)} = w^{(i)}\,v_{s,d}^{(i-1)} + c_1^{(i)}\,\left(p_{s,d}^{(i-1)}-x_{s,d}^{(i-1)}\right)+c_2^{(i)}\,\left(p_{g_s,d}^{(i-1)}-x_{s,d}^{(i-1)}\right)
\end{align}
with inertia $w^{(i)}$ and random coefficients $c_{1/2}^{(i)}\in[0,C_{1/2}^{(i)}]$. We initialize these dynamic parameters with ${w^{(1)}=0.42}$ and ${C_{1/2}^{(1)}=1.55}$, respectively, as recommended in Ref.~\cite{Liu2014}, see also Ref.~\cite{Bonyadi2017}, and optionally modify them according to an adaptive schedule during the course of optimization. Once we have moved the whole swarm, we enforce all constraints, which poses no problem since the position update is random anyway, and then make the velocities consistent with the constraints. For example, in the case of Eq.~(\ref{E}), we enforce the constraint ${\sum_{i=1}^Ln_i=N}$ of Eq.~(\ref{OccNumConstraint2}) as illustrated with Fig.~\ref{FigurePSO}b. There, $\dens{n}^{\mathrm{req}}$ is first rescaled to ${\tilde{\dens{n}}=\frac{N}{N^{\mathrm{req}}}\dens{n}^{\mathrm{req}}}$ if ${N^{\mathrm{req}}=\sum_{i=1}^Ln_i^{\mathrm{req}}\not=N}$. Then, we obtain $\dens{n}$ by adding a rescaled ${\vec \sigma=\tilde{\dens{n}}-\dens{n}_{\mathrm c}}$, which is parallel to the constraining line, to ${\dens{n}_{\mathrm c}=\frac{N}{L}(1,1)}$, which points to the center of the intersection of the constraining line with the square that encodes ${0\le n_i\le2}$. We rescale $\vec \sigma$ using the maximal excess ${|\delta \sigma|=\mathrm{max}_{i}\big\{\mathrm{max}\{n_i-2,-n_i\}\big\}}$, here incidentally realized by $|\delta \sigma_2|$, among all participation numbers beyond their allowed values in $[0,2]$. The generalization to larger $L$ is straightforward. As an alternative to rescaling $\dens{n}^{\mathrm{req}}$ to $\tilde{\dens{n}}$ on the constraining (hyper-)plane defined by ${\sum_{i=1}^Ln_i=N}$, we may \textit{project} onto the plane, which generally results in a different $\tilde{\dens{n}}$ on the plane. In fact, we alternate between both options, rescaling and projection, in a random fashion. Note that the PSO never requests participation numbers outside $[0,2]$ since the participation numbers are constructed as ${n_i=1+\cos\big(\theta_i\big)}$, but some $n_i$ can exceed $2$ after rescaling if ${N^{\mathrm{req}}<N}$, and the projection can even result in negative excess $\delta \sigma$. After the constraints are enforced, we calculate $f\left(\vec x_s^{(i)}\right)$ for all $s$ and prepare to move the swarm in the next iteration. We monitor the swarm by collecting the best function values ${\vec f_g = \left(f_g^{(i)},\dots, f_g^{(i-\lambda+1)}\right)}$ of the $\lambda$ most recent iterations. We declare convergence and terminate the PSO run once the current variance $C(i)$ of $\vec f_g$ falls below a chosen target variance $T$; our default settings are ${\lambda=\mathrm{min}(30,D)}$ and ${T=10^{-12}D}$. Since PSO is stochastic and can, despite all our efforts, get stuck in local optima, it is prudent to judge the quality of an alleged global optimum with the aid of a histogram of the optima from several (e.g., ${10\mbox{ to }1000}$) PSO runs, see Fig.~\ref{FigurePSO}c.

We designed several adaptations to this general scheme to aid the exploratory capability of the swarm in the initial phase of each run, the convergence in the final phase, and the swarm's ability to avoid premature convergence to local optima in the intermediate phase. We choose the number of elements in each group $G_s$ to about $S/4$ and randomly re-initialize this grouping, together with the associated inter-particle links, if the current iteration does not improve upon the best function value encountered during the current PSO run. In other words, there are effectively four concurrent swarms with populations that interchange their members in an adaptive fashion. In our experience, ${40\lesssim S_0\lesssim 10D}$ is a suitable range for the initial swarm size ${S=S_0}$, with a problem-dependent trade-off between swarm size and required iterations/runs. Our PSO code is parallelized with openMP, and each swarm particle is processed in a separate thread. With decreasing difference between current and target variance, we decrease $S$ down to $\mathrm{max}(10,t)$, where $t$ is the number of available parallel-processing threads. The worst particles (in terms of their personal best) are thereby discarded according to the schedule ${S(i)=\mathrm{min}\left\{S_0,\mathrm{max}\left[t,t+S_0\,\left(1-\mathrm{log}_{10}C(i)/\mathrm{log}_{10}T\right)\right]\right\}}$ at iteration $i$. This procedure removes inferior and/or superfluous particles as long as the variance drops and hastens the convergence in the later optimization stages.

Furthermore, much has been said about the curse of dimensionality, viz., the exponential growth of the search space with $D$, which lies at the heart of so many real-world problems, not least the quantum many-body problem. But there is also a blessing of dimensionality, namely that reducing the search interval in a \textit{single} dimension by $50\%$ removes half of the \textit{total} search space, irrespective of $D$. We therefore implemented an adaptive search space, where search space intervals $I_d$ shrink in all those dimensions $d$, for which the coordinate of a newly encountered global best lies close to the center of $I_d$. In turn, we shift $I_d$, to the extent the constraints allow, whenever the $d$th coordinate of a new global best comes close to a boundary of $I_d$. Optionally, the so adapted search space can be inherited by subsequent runs. If active, this procedure eases the computational load by focusing on more promising regions---at the expense of potentially sacrificing superior regions that can be reached only after sustained uphill exploration; but in practice, we find essentially the same high-quality optima with or without an adaptive search space, just more efficiently in the former case.

By default, we make the dynamic PSO parameters $w$ and $C_{1/2}$ self-adapting: every time a significantly better global best is found, we draw a new, say, $w$ from a Gaussian distribution with variance of $0.0025$, centered at $w_{\mathrm{c}}$, which is itself updated according to $w_{\mathrm{c}}\rightarrow I*w_{\mathrm{c}}+(1-I)*\langle w\rangle$, with inertia ${I=0.8}$ and the mean $\langle w\rangle$ of previously successful (in finding a significantly better global best) parameters $w$. Each particle holds its own parameters $w$ and $C_{1/2}$, independent from the other particles. Hence, at any given iteration, the swarm's individuals span a self-adapting distribution of search capabilities.

Finally, we note our unsuccessful attempts to improve the PSO performance through self-adaption of the PSO hyperparameters ($w$, $C_1$, $C_2$, $S$, $\lambda$, $T$, etc.) by adding them to the search space $X$ or through `fuzzy self-tuning' \cite{Nobile2018}. Also of little effect was our attempt to counter premature convergence by maintaining diversity among the swarm particles in the intermediate optimization stage: we replaced the worst fraction $\sigma$ of particles (five times per run on average) with randomly re-initialized ones. We reckoned this to have little effect in the very early exploratory stages, which are reminiscent of random search anyway, or in the final stages, where the pull of many near-optimal particles dominates the swarm behavior. Convergence can also be accelerated through elitism, where the global best particle informs all others with a finite probability. However, we found it expedient to deactivate elitism by default to better escape local optima.

\subsection{Genetic algorithm}

\enlargethispage{\baselineskip}

Genetic algorithms (GAs) are a class of evolutionary algorithms that resemble Darwinian evolution. The principles of natural selection are mimicked through crossover, mutation, and selection operators that act on a population of `chromosomes', each being a vector of variables (`genes') from the search space of the objective function. While many different flavors of GAs have been studied and countless heuristics proposed for augmenting these GAs \cite{Slowik2020}, the following procedure is typically iterated: given a number of `parents' whose fitness (viz., objective function value) has been determined, two parents are chosen for breeding a `child' by exchanging genes according to a crossover rate, followed by randomly altering genes of the child according to a mutation rate and enforcing problem-specific constraints on the genes. In a subsequent selection process, children (or random invaders according to an invasion rate) may then replace parents based on a fitness comparison. For multi-modal deceptive objective functions, this selection pressure commonly homogenizes the population too quickly \cite{Hutter2006,Park2010}, with a large part of the population representing the same local optimum---although mutations and invasions allow, in principle, the exploration of the whole search space. We thus built our genetic algorithm optimization (GAO), which we used to validate the results obtained with PSO, by augmenting an otherwise prototypical GA with three heuristics. First, we select new generations of parents akin to the fitness uniform selection scheme \cite{Hutter2006}, which guarantees a high diversity among chromosomes throughout the entire evolution, though we skew the selection toward fitter parents and always preserve the fittest chromosome. Second, we let many small subpopulations evolve in parallel, with inter-population breeding that begins with neighboring populations according to a dispersal parameter and extends to all populations toward the end of a chosen maximum number of generations. Finally, we also shrink the population sizes to accelerate the convergence toward the end of the optimization, analogous to our procedure for PSO that is controlled by the history of encountered objective function values.

\section{\label{AppendixInteractionTensorElements}Interaction tensor elements}

The interaction tensor elements $I_{abcd}$ of Eq.~(\ref{Tabcd}) for the specific physical systems simulated in this work are available in analytical form (except for the results on relativistic atoms presented in Sec.~\ref{RelativisticAtoms}), see \ref{AppendixContactInteraction1D}--\ref{AppendixCoulombInteraction3D} below. For the small-scale systems that are addressed in this work with objective-function-based optimizers, it is expedient to precompute the elements $I_{abcd}$ and store them in compact form by making use of their system-dependent symmetries.

\subsection{\label{AppendixContactInteraction1D}Contact interaction (1D)}

We compute the interaction tensor elements $I_{ a  b  c  d}$ of Eq.~(\ref{Tabcd}) for unpolarized contact-interacting spin-$\half$ fermions of mass $m$ in a one-dimensional harmonic oscillator of frequency $\omega$ with the aid of the generating function
\begin{align}
    f_K(x,y)=\sum_{a=0}^{\infty} \sum_{b=0}^{\infty}\frac{x^{a}}{\sqrt{a!}}\frac{y^{b}}{\sqrt{b!}}\langle a |\Exp{\I K(\mathfrak{a} + \mathfrak{a}^{\dagger})}| b \rangle = \Exp{xy} \Exp{\I K(x+y)}  \Exp{-\frac{K^{2}}{2}}\,,
\end{align}
where ${K=k \sqrt{\frac{\hbar}{2m\omega}}}$ and ${\mathfrak{a}+\mathfrak{a}^{\dagger}=\sqrt{\frac{2m\omega}{\hbar}}R}$, such that
\begin{align}
\langle a |\Exp{\I k R}| b \rangle=\langle a |\Exp{\I K(\mathfrak{a} + \mathfrak{a}^{\dagger})}| b \rangle=\frac{1}{\sqrt{a!}}\frac{1}{\sqrt{b!}}\Big(\frac{\partial}{\partial x}\Big)^{a}\Big(\frac{\partial}{\partial y}\Big)^{b} f_K(x,y) \Bigg|_{x=y=0}=\Exp{\frac{-K^{2}}{2}}\frac{\sqrt{a!}}{\sqrt{b!}}(\I K)^{b-a}L_a^{(b-a)}(K^2)\,,
\end{align}
which is symmetric in the nonnegative integers $a$ and $b$; hence, $b \geq a$ without loss of generality. For the contact interaction in Eq.~(\ref{VintContact}),
we have ${u(k)=\mathpzc{c}}$ in Eq.~(\ref{Tabcd}). Then, expressing the generalized Laguerre polynomials as ${L_n^{\alpha}(x)=\sum_{j=0}^{n}\binom{n+\alpha}{n-j}\frac{(-x)^{j}}{j!}}$, while setting ${n= a}$ (${n= d}$) and ${\alpha= b- a}$ (${\alpha'= c- d}$), we write
\begin{align}
\sqrt{\frac{2\hbar}{m\omega}}\,\pi\,I_{ a  b  c  d}&= \int \d K \,\mathpzc{c}\,\langle  a |\Exp{\I K(\mathfrak{a} + \mathfrak{a}^{\dagger})} |  b \rangle \langle  c |\Exp{-\I K(\mathfrak{a} + \mathfrak{a}^{\dagger})} |  d \rangle\nn\\
&=\int \d K \,\mathpzc{c}\, \Exp{-K^{2}}\sqrt{\frac{ a! d!}{ b! c!}}(\I K)^{ b- a}(-\I K)^{ c- d}L_{ a}^{\alpha}(K^2)L_{ d}^{\alpha'}(K^2)\nn\\
&=\int \d K \,\mathpzc{c}\, \Exp{-K^{2}}\sqrt{\frac{ a! d!}{ b! c!}}\sum_{j=0}^{ a}\sum_{j'=0}^{ d}\binom{ b}{ a-j}\binom{ c}{ d-j'}\frac{(-K^{2})^{G}}{j!j'!}(-1)^{ b- a}\,,\label{TabcdContactInteraction}
\end{align}
where ${G=\frac12( b- a+  c- d)+j+j'}$. Since ${I_{ a  b  c  d}=I_{ b a   d c }=I_{dcba}=I_{cdba}}$, it suffices to compute $I_{ a  b  c  d}$ for index sets with $ a \leq  b \leq c$ and $ c \geq  d$, for which Eq.~(\ref{TabcdContactInteraction}) reduces to
\begin{align}\label{TabcdContactInteractionSum}
I_{ a  b  c  d}=
\begin{cases}
\mathpzc{c} \sqrt{\frac{m \omega}{\hbar}}\sum_{j=0}^{ a}\sum_{j'=0}^{ d}\frac{1}{\sqrt{2\pi}}\frac{(-1)^{ b- a}\sqrt{ a! b! c! d!}(-\frac{1}{4})^{G}(2G)!}{( b- a+j)!( a-j)!( c- d+j')!( d-j')!j!j'!G!} & ,\; a+b+c+d\;\mbox{ even}\\
0 &,\; a+b+c+d\;\mbox{ odd}
\end{cases}\,.
\end{align}

As an alternative to Eq.~(\ref{TabcdContactInteractionSum}), we derive a recursion relation for
\begin{align}\label{HabcdContactalternative}
\mathcal{H}_{abcd} = \frac{\mathpzc{c}}{\sqrt{8 \pi} \Ell} \frac{ J {\pqty{a,\, b,\, c,\, d}}}{\sqrt{2^{a} a! 2^{b} b! 2^{c} c! 2^{d} d!}}
\end{align}
in Eq.~(\ref{eq:D3}). Here,
\begin{align}
J {\pqty{a,\, b,\, c,\, d}} = \sqrt{\frac{2}{\pi}} \int \dd{x}\, \Exp{-2x^2}\, \Herm{a}{x}\, \Herm{b}{x}\, \Herm{c}{x} \,\Herm{d}{x}\, ,
\end{align}
with the Hermite polynomial $\Herm{n}{x}$ of order ${n}$, obeys the relation
\begin{align} \label{eq:GenFuncI}
	\sum_{a,\, b,\, c,\, d} \frac{z_1^{a}}{a!} \frac{z_2^{b}}{b!} \frac{z_3^{c}}{c!} \frac{z_4^{d}}{d!} J {\pqty{a,\, b,\, c,\, d}} = \exp\left[\frac{1}{2} {\pqty{z_1 + z_2 + z_3 + z_4}}^2 - {\pqty{z_1^2 + z_2^2 +z_3^2 + z_4^2}} \right] \,,
\end{align}
obtained from applying the generating function of the Hermite polynomials four times. Operating with ${z_1 \pdv{z_2}}$ on Eq.~\eqref{eq:GenFuncI}, we find
\begin{align}
J {\pqty{a-1,\, b+1,\, c,\, d}}=
\begin{cases}
\;0 & ,\; a + b + c + d \;\text{ odd}\\
\;\mbox{\parbox{20em}{${\pqty{a-1}}\, J {\pqty{a-2,\, b,\, c,\, d}} - b\, J {\pqty{a-1,\, b-1,\, c,\, d}}\\ + c\, J {\pqty{a-1,\, b,\, c-1,\, d}} + d\, J {\pqty{a-1,\, b,\, c,\, d-1}}$ }} & ,\; a + b + c + d \;\text{ even}\\
\end{cases}
\end{align}
with initial values $J {\pqty{a,\, 0,\, 0,\, 0}} = {\pqty{-\frac{1}{2}}}^{a/2} \frac{a!}{(a/2)!}\,[(a+1)\mbox{ mod }2]$. Following Ref.~\cite{Na+1:17}, we may also write
\begin{equation}
	J {\pqty{a,\, b,\, c,\, d}} = \sum_{q=0}^{\Min{a+b,c+d}} \delta_{(a+b),(c+d)}\,\delta_{(a+b),(q\text{ mod }2)}\, \alpha_q \pqty{a,\, b} \alpha_q \pqty{c,\, d} 2^q q! \,,
\end{equation}
where $\alpha_q \pqty{a,\, b} = \sum_{n=\Max{0,q-a}}^{\Min{b,q}} \pqty{-1}^{b-n+ \frac{a+b-q}{2}} \left[2^{\frac{a+b}{2}} q! {\pqty{\frac{a+b-q}{2}}}! \right]^{-1} \binom{q}{n} \binom{a+b-q}{b-n} $.

The accurate numerical evaluation of the closed formulae in Eqs.~(\ref{TabcdContactInteractionSum}) and (\ref{HabcdContactalternative}) beyond $L\approx15$ energy levels require high-precision arithmetic. Alternatively and as a means of validation, the tensor elements $I_{ a  b  c  d}$ can be tabulated using double-precision arithmetic by directly evaluating the integral in Eq.~(\ref{TabcdContactInteraction})---we produced and confirmed all tensor elements with indices up to ${L=100}$ using an adaptive bisection algorithm with Boole quadrature.

\subsection{\label{AppendixHarmonicInteraction1D}Harmonic interaction (1D)}

With the oscillator eigenfunctions $\bk{x}{ a}=\psi_{ a}(x)$ and ${V_{\mathrm{int}}(x_j-x_k)=\frac{\mathcal{E}}{\mathcal{L}^2}\tilde\beta\,(x_j-x_k)^2=\beta\,(x_j-x_k)^2}$ from Eq.~(\ref{Vintharmonic}), we get
\begin{align}
I_{ a b c d}=\mathcal{E}\tilde\beta\,\int\d\tilde x\,\d\tilde x'\,(\tilde x-\tilde x')^2\,\mathcal{L}^2\psi_{ a}(x)\,\psi_{ b}(x)\,\psi_{ c}(x')\,\psi_{ d}(x')\,,
\end{align}
with $\tilde x=x/\mathcal{L}$, for $\trace\left(\rho_{\mathrm{mp}}H_{\mathrm{int}}\right)$ in Eq.~(\ref{eq:D3}). Here, we explicitly exhibit the harmonic oscillator units of energy ${\mathcal{E}=\hbar\omega}$ and length ${\mathcal{L}=\sqrt{\hbar/(m\omega)}}$, with particle mass $m$ and oscillator frequency $\omega$, i.e., the units of the noninteracting system (${\alpha=1}$)). Then, we write
\begin{align}
\frac{1}{\mathcal{E}\tilde\beta}I_{abcd}&=\left[J_{ab}^{(2)}\delta_{cd}-J_{ab}^{(1)}J_{cd}^{(1)}\right]+[a\leftrightarrow d\; \&\; b\leftrightarrow c]\,,
\end{align}
where $J_{ab}^{(\nu)}=\int\d\tilde x\, \tilde x^{\nu}\,\tilde\psi_{a}(\tilde x)\tilde\psi_{b}(\tilde x)$, with $\tilde\psi_{j}(\tilde x)=\sqrt{\mathcal{L}}\psi_{j}(x)=\sqrt{\mathcal{L}}\psi_{j}(\mathcal{L} \tilde x)$.

The recurrence relation
\begin{align}
H_{j+1}(\tilde x)=2\tilde x\,H_j(\tilde x)-2j\,H_{j-1}(\tilde x)
\end{align}
for the Hermite polynomials implies the recurrence relation
\begin{align}
\tilde x\,\tilde \psi_{j}(x)=\sqrt{\frac{j+1}{2}}\tilde \psi_{j+1}(\tilde x)+\sqrt{\frac{j}{2}}\tilde \psi_{j-1}(\tilde x)
\end{align}
for the Hermite functions $\tilde \psi_{j}(\tilde x)=(2^j\,j!)^{-1/2}\pi^{-1/4}\exp\left(-\tilde x^2/2\right)H_j(\tilde x)$. Hence, with ${M=\mbox{max}(a,b)}$ and ${m=\mbox{min}(a,b)}$, we get
\begin{align}
J_{ab}^{(1)}=
\begin{cases}
0 & ,\; \mbox{both } a\; \&\; b \mbox{ even or odd (`same parity')}\\
\int\d\tilde x\, \tilde x\,\tilde \psi_{M}(\tilde x)\tilde \psi_{m}(\tilde x)=\sqrt{\frac{M}{2}}\delta_{m,M-1} & ,\; a\; \&\; b \mbox{ different parity }
\end{cases}
\end{align}
and
\begin{align}
J_{ab}^{(2)}=
\begin{cases}\label{J2}
0 & ,\; a\; \&\; b \mbox{ different parity}\\
1/2 & ,\; a=b=0\\
3/2 & ,\; a=b=1\\
\frac12(2a+1) & ,\; a=b>1\\
\frac12\sqrt{M(M-1)}\,\delta_{M-2,m} & ,\; M\; \&\; m \mbox{ same parity and } M\ge m+2
\end{cases},
\end{align}
where the last two cases of (\ref{J2}) follow from
\begin{align}
\tilde x^2\tilde \psi_M(\tilde x)=\frac12\left(\sqrt{(M+1)(M+2)}\tilde \psi_{M+2}(\tilde x)+(2M+1)\tilde \psi_{M}(\tilde x)+\sqrt{M(M-1)}\tilde \psi_{M-2}(\tilde x)\right),
\end{align}
and the three central cases of (\ref{J2}) are summarized by $\frac12(2a+1)$ for ${a=b}$.\\

Since $n(-x)=n(x)$ for the ground-state density $n(x)$, we write
\begin{align}
\left.\trace\left(\rho_{\mathrm{mp}}H_{\mathrm{int}}\right)\right|_{\mbox{direct part}}=\frac12\int\d x\,\d x'\,\beta\,(x-x')^2\,n(x)\,n(x')=\int\d x\,\frac12\,2\beta N\,x^2\,n(x)\,,
\end{align}
such that the total energy (with only the direct part of the interaction) for any $\alpha$ is
\begin{align}
E_{\mathrm d}^{(\alpha)}=E_{\mathrm{kin}}+\int\d x\,\frac12 (m\omega^2+2\beta N)\,x^2\,n(x)\,.
\end{align}
With $\beta=\frac{\mathcal{E}}{\mathcal{L}^2}\tilde\beta$, we have
\begin{align}
m\omega^2+2\beta N=\hbar\omega\frac{m\omega}{\hbar}+2\frac{\mathcal{E}}{\mathcal{L}^2}\tilde\beta N=\frac{\mathcal{E}}{\mathcal{L}^2}\underset{=\alpha}{\underbrace{(1+2\tilde\beta N)}}=\hbar\omega\sqrt{\alpha}\frac{m\omega\sqrt{\alpha}}{\hbar}\,.
\end{align}
Hence, 
\begin{align}\label{Ealpha}
E_{\mathrm d}^{(\alpha)}=E_{\mathrm{kin}}+\mathcal{E}\int\d\tilde x\,\frac12\alpha\,\tilde x^2\,\tilde n(\tilde x)\,,
\end{align}
where $\tilde n(\tilde x)=\mathcal{L} n(\tilde x)$. That is, (\ref{Ealpha}) is the energy of a noninteracting harmonic oscillator with frequency $\omega\sqrt{\alpha}$ (in units of the oscillator with frequency $\omega$) and can be directly utilized in DFT calculations---e.g., for comparison between 1pEx-DFT and DPFT. The virial theorem then implies $E_{\mathrm d}^{(\alpha)}=\alpha\,E^{(1)}_{\mathrm d}$, and the $\alpha$-dependent dimensionless eigenfunctions
\begin{align}
\tilde \psi_j^{(\alpha)}(\tilde x)=(2^j\,j!)^{-1/2}\pi^{-1/4}\alpha^{1/8}\exp\left(-\frac{\sqrt{\alpha}}{2}\tilde x^2\right)H_j\left(\alpha^{1/4}\tilde x\right)
\end{align}
yield (for even $N$) the exact dimensionless ground-state density
\begin{align}
\tilde n(\tilde x)=2\sum_{j=0}^{\frac{N}{2}-1}|\tilde \psi_j^{(\alpha)}(\tilde x)|^2
\end{align}
of the interacting system.

\subsection{\label{AppendixCoulombInteraction3D}Coulomb interaction (3D)}

\subsubsection{\label{AppendixNonrelativisticAtoms}Nonrelativistic atoms}

We evaluate the interaction energy in Eq.~(\ref{eq:D3}) for atoms by calculating the tensor element $I_{abcd}$ of Eq.~(\ref{Tabcd}). First, we consider nonrelativistic atoms as a blueprint for and a comparison with the relativistic case. Therefore, $I_{abcd}$ takes the nonrelativistic hydrogenic wave functions
\begin{align}\label{HydrogenicWaveFunction}
\psi_a(\vec r) = \psi_{n_a,l_a,m_a}(r,\theta,\phi) = \gamma_a\,R_a(r)\,Y_{l_a,m_a}(\theta,\phi)
\end{align}
(and accordingly for the orbital indices $b,c,d$), where the orbital index ${a=1,2,\dots}$ combines the three hydrogenic quantum numbers ${\{n_a=1,2,\dots;l_a=0,1,\dots,n_a-1;m_a=-l_a,-l_a+1,\dots,l_a\}}$ in row-major order (${\dots,\,a=5\leftrightarrow\{n_a=2;l_a=1;m_a=1\},\,a=6\leftrightarrow\{n_a=3;l_a=0;m_a=0\},\dots}$). We choose the phase factors $\gamma_a=1$. The radial wave functions are
\begin{align}
R_a(r) = \sqrt{\left(\frac{2Z}{n_a}\right)^3\frac{(n_a-l_a-1)!}{2n_a\,(n_a+l_a)!}}\,\left(\frac{2Zr}{n_a}\right)^{l_a}\exp\left(\frac{-Zr}{n_a}\right)\,L_{n_a-l_a-1}^{2l_a+1}\left(\frac{2Zr}{n_a}\right)\,,
\end{align}
with the generalized Laguerre polynomials
\begin{align}\label{Laguerre}
L_{n}^{\alpha}(x)=\sum_{l=0}^n\frac{(\alpha+l+1)_{n-l}}{(n-l)!\,l!}(-x)^l\,,
\end{align}
where ${(z)_n=z\,(z+1)\dots(z+n-1)}$.

\begingroup
\allowdisplaybreaks

Then, we proceed along the lines of Ref.~\cite{Condon1935} with the help of the Laplace expansion
\begin{align}
\frac{1}{|\vec r-\vec r'|}=\sum_{k=0}^\infty\frac{4\pi}{2k+1}\sum_{m=-k}^k(-1)^m\frac{r_<^k}{r_>^{k+1}}\,Y_{k,-m}(\theta,\phi)\,Y_{k,m}(\theta',\phi'),
\end{align}
where $\{r,\theta,\phi\}$ $\big(\{r',\theta',\phi'\}\big)$ are the spherical coordinates of $\vec r$ ($\vec r'$), ${r_<=\Min[]{r,r'}}$, ${r_>=\Max[]{r,r'}}$, and $Y_{k,m}(\theta,\phi)$ are the spherical harmonics with the phase convention used in \cite{NISTHandbook}. Using the relation ${Y_{k,m}^*(\theta,\phi)=(-1)^m\,Y_{k,-m}(\theta,\phi)}$, we write
\begin{align}\label{TabcdSum}
I_{abcd}=\gamma_a\gamma_b^*\gamma_c\gamma_d^*\sum_{k=0}^\infty\frac{4\pi}{2k+1}\sum_{m=-k}^k(-1)^{m+m_b+m_d}\,R^k(ac,bd)\,\tilde{S}^{l_a,l_b,k}_{m_a,-m_b,-m}\,\tilde{S}^{l_c,l_d,k}_{m_c,-m_d,m}
\end{align}
with the dimensionless radial integrals
\begin{align}
R^k(ab,cd) = Z\,R_H^k(ab,cd) = \int_0^\infty\d r\d r'\,r^2\,r'^2\frac{r_<^k}{r_>^{k+1}}\,R_a(r)\,R_b(r')\,R_c(r)\,R_d(r')\,,
\end{align}
where $R_H^k(ab,cd)$ is independent of $Z$, and the Gaunt coefficients \cite{NISTHandbook}
\begin{align}
\tilde{S}^{j_1,j_2,j_3}_{m_1,m_2,m_3} = \sqrt{\frac{(2j_1+1)(2j_2+1)(2j_3+1)}{4\pi}}\,\left(\begin{array}{ccc} j_1 & j_2 & j_3\\ 0 & 0 & 0\end{array}\right)\,\left(\begin{array}{ccc} j_1 & j_2 & j_3\\ m_1 & m_2 & m_3\end{array}\right) =: \sqrt{\frac{(2j_3+1)}{4\pi}} S^{j_1,j_2,j_3}_{m_1,m_2,m_3}
\end{align}
that include a product of two 3j-symbols and, hence, can be nonzero only if ${|j_1-j_2|\le j_3\le j_1+j_2}$, ${m_1+m_2+m_3=0}$, and ${j_1+j_2+j_3}$ even. That is, Eq.~(\ref{TabcdSum}) reduces to
\begin{align}\label{TabcdFinal}
I_{abcd}=\sum_{k=k_{\mathrm{min}}}^{k_{\mathrm{max}}} \gamma_{abcd}(k)\,R_H^k(ac,bd)\,S^{l_a,l_b,k}_{m_a,-m_b,m_b-m_a}\,S^{l_c,l_d,k}_{m_c,-m_d,m_d-m_c}
\end{align}
with ${k_{\mathrm{min}}=\Max[]{|l_a-l_b|,|l_c-l_d|}}$, ${k_{\mathrm{max}}=\Min[]{l_a+l_b,l_c+l_d}}$, and
\begin{align}
\gamma_{abcd}(k)=\,Z\,\gamma_a\gamma_b^*\gamma_c\gamma_d^*(-1)^{m_a+m_d}\,\delta_{m_a-m_b,m_d-m_c}\,[(l_a+l_b+k+1)\mbox{ mod }2]\,[(l_c+l_d+k+1)\mbox{ mod }2]\,.
\end{align}
We precompute the radial integrals
\begin{align}
R_H^k(ab,cd) &= G_H^k(ac,bd)+G_H^k(bd,ac)\label{RHk}
\end{align}
after expanding the generalized Laguerre polynomials according to Eq.~(\ref{Laguerre}), with
\begin{align}
G_H^k(ac,bd) &= \frac{1}{Z}\int_0^\infty\d r\,r^{2-k-1}R_a(r)R_c(r) \int_0^r\d r'\,r'^{2+k}R_b(r')R_d(r')\nn\\
&=\tau_a\tau_b\tau_c\tau_d\sum_{ a_a=0}^{n_a-l_a-1}\sigma_a( a_a)\sum_{ a_b=0}^{n_b-l_b-1}\sigma_b( a_b)\sum_{ a_c=0}^{n_c-l_c-1}\sigma_c( a_c)\sum_{ a_d=0}^{n_d-l_d-1}\sigma_d( a_d)\nn\\
&\quad\times g^k(1-k+ a_a+l_a+ a_c+l_c,\frac{1}{n_a}+\frac{1}{n_c};2+k+ a_b+l_b+ a_d+l_d,\frac{1}{n_b}+\frac{1}{n_d})\,,\label{GHk}\\
\tau_a &=\sqrt{\left(\frac{2}{n_a}\right)^3\frac{(n_a-l_a-1)!}{2n_a\,(n_a+l_a)!}}\,\left(\frac{2}{n_a}\right)^{l_a}\,,\\
\sigma_a( a_a)&=\frac{(2l_a+2+ a_a)_{n_a-l_a-1- a_a}}{(n_a-l_a-1- a_a)!\, a_a!}\,\left(\frac{-2}{n_a}\right)^{ a_a}\,,\\
g^k(p,\alpha;q,\beta) &= \int_0^\infty\d\zeta\,\zeta^p\,\mathrm{e}^{-\alpha\zeta}\int_0^\zeta\d\zeta'\,\zeta'^q\,\mathrm{e}^{-\beta\zeta'}\nn\\
&= \frac{p!}{\beta^{q+2}}\left[\frac{\beta\,q!}{\alpha^{p+1}}-\frac{(p+q+1)!}{\beta^p\,(p+1)!}\,{_2F_1}\left(\begin{array}{c}p+1,p+q+2\\ p+2\end{array};\,-\frac{\alpha}{\beta}\right)\right]\,,\label{gk}
\end{align}
and the generalized hypergeometric function ${{_2F_1}\left(\begin{array}{c}u,v\\ w\end{array};\,z\right)=\sum_{k=0}^\infty\frac{(u)_k\,(v)_k}{(w)_k}\frac{z^k}{k!}}$. Equation~(\ref{gk}) derives from changing the integration variable $r$ in Eq.~(\ref{GHk}) to ${\zeta=Zr}$. 
The numerical values we obtain for Eq.~(\ref{RHk}) coincide with those reported in Refs.~\cite{Joshi1971,Butler1971,Butler1971Erratum}.

\endgroup

The core Hamiltonian $H_{\mathrm{1p}}$ of atoms and ions has scattering states, which we disregarded when using only the bound states defined in Eq.~(\ref{HydrogenicWaveFunction}) for generating the 1pEx binding energies in Sec.~\ref{RelativisticAtoms}; cf.~the discussion around Eq.~(\ref{eq:A3}). This omission is responsible for the discrepancies between the energies from 1pEx-DFT and HF in the case of ${N=2}$ electrons, see Fig.~\ref{RelDev_1pEx_HF_NIST}. Indeed, the accumulated overlap $\sum_{s=1}^\infty|\bk{\phi}{s}|^2$ between the HF ground-state orbital $\ket{\phi}$, taken from Ref.~\cite{King2018}, and the hydrogenic s-states $\ket{s}$ is less than one: approximately 0.994365 for H$^{-}$, 0.994945 for He, 0.999156 for C$^{4+}$, and 0.999898 for Ar$^{16+}$. That is, in these cases a small part of $\ket{\phi}$ is composed of scattering states.

\subsubsection{\label{AppendixRelativisticAtoms}Relativistic atoms}

We reduce the program of the exact two-component quasi-relativistic theory (X2C) to a one-component (viz., spin-free) quasi-relativistic theory (sf-X2C) along the lines of Ref.~\cite{Kutzelnigg2005}. In summary, the resulting relativistic hydrogenic states $\psi_k(\vec r)=\bk{\vec r}{k}=\sum_{\mu=1}^M C_{\mu k}\,g_\mu(\vec r)$, here expanded in the nonrelativistic hydrogenic states $\{g_\mu\}$ as given in Eq.~(\ref{HydrogenicWaveFunction}), yield the associated relativistic interaction tensor elements
\begin{align}\label{RelTabcd}
I_{abcd}=\sum_{\kappa,\lambda,\mu,\nu=1}^M C_{\kappa a}\,C_{\lambda b}^*\,C_{\mu c}\,C_{\nu d}^*\,I^{\mathrm{NR}}_{\kappa\lambda\mu\nu}\,.
\end{align}
Here, $I^{\mathrm{NR}}_{\kappa\lambda\mu\nu}$ are the nonrelativistic interaction tensor elements given in Eq.~(\ref{TabcdFinal}), and $M$ controls the quality of this expansion; we choose ${M=L}$ for simplicity. Although other bases such as Slater-orbitals are thinkable, we opt for the nonrelativistic hydrogenic states in order to keep $M$ small, because the values $\left\{I^{\mathrm{NR}}_{\kappa\lambda\mu\nu}\right\}$ are already available in this case, and for convenient consistency checks with light atoms, where ${I_{abcd}\approx I^{\mathrm{NR}}_{abcd}}$.

Defining the ${M\times M}$-dimensional (unitless) matrices
\begin{align}
S_{\mu\nu}&=\bk{g_\mu}{g_\nu}=\delta_{\mu\nu}\,,\\
T_{\mu\nu}&=\frac{1}{\mbox{Ha}}\bok{g_\mu}{\frac{\vec P^2}{2m_{\mathrm{e}}}}{g_\nu}\,,\label{X2C_T}\\
V_{\mu\nu}&=\frac{1}{\mbox{Ha}}\bok{g_\mu}{V_\mathrm{ext}(\vec R)}{g_\nu}\,,\\
\intertext{and}
U_{\mu\nu}&=\frac{1}{m_{\mathrm{e}}\,\mbox{Ha}^2}\bok{g_\mu}{\sum_{j=1}^3 P_j\,V_\mathrm{ext}(\vec R)\,P_j}{g_\nu}\,,\label{X2C_U}
\end{align}
we find the coefficients $C_{\mu k}$ as the matrix elements of ${C=\tilde{S}^{\half}\,A}$. Here, ${\tilde{S}=S+\frac{1}{2m_{\mathrm{e}}} X^{\dagger}T^{-1} X}$ is a modified metric, and we adopt atomic units of Hartree (Ha), Bohr radius ($a_0$), and electron rest mass ($m_{\mathrm{e}}$), see also Sec.~\ref{RelativisticAtoms}. The matrix ${X=2\alpha\sqrt{m_{\mathrm{e}}\,\mbox{Ha}}\,TBA^{-1}}$, where $\alpha$ is the fine structure constant, is built from the nonrelativistic kinetic energy matrix $T$ as given in Eq.~(\ref{X2C_T}) and from the matrices $A$ and $B$, whose elements are found by solving the generalized eigenvalue problem that is presented by the one-particle Dirac equation in (block-)matrix form, with energies shifted by the electron rest-mass energy:
\begin{align}
\left(
\begin{array}{cc}
 V & 2T\\
 2T & \alpha^2U-4T
\end{array}
\right)
\left(
\begin{array}{c}
 \vec a_k\\
 \vec b_k
\end{array}
\right)
=
\frac{E_k}{\mbox{Ha}}
\left(
\begin{array}{cc}
 S & 0\\
 0 & 2\alpha^2T
\end{array}
\right)
\left(
\begin{array}{c}
 \vec a_k\\
 \vec b_k
\end{array}
\right),\mbox{ for all }k=1,\dots,M\,.\label{modifiedDE}
\end{align}
Here, the (unitless) vector ${\vec a_k=(A_{1k},A_{2k},...,A_{Mk})}$, i.e., the $k$th column of $A$, defines the $\{g_\mu\}$-expansion of the so-called large component $\varphi_k(\vec r)=\sum_{\mu=1}^M A_{\mu k}\,g_\mu(\vec r)$ of the Dirac spinor that solves the Dirac equation for eigenenergy $E_k/\mbox{Ha}$. While $\varphi_k$ is a two-component spinor in full four-component Dirac theory, it is a scalar quantity in the sf-X2C employed here. The $M$ electronic eigenstates among the $2M$ eigenstates of Eq.~(\ref{modifiedDE}) are associated with energies ${E_k>-m_\mathrm{e}c^2}$ \cite{Kutzelnigg2005}, where $c$ is the speed of light, i.e., ${E_k/\mbox{Ha}>-1/\alpha^2}$. The key element of X2C is the matrix~$X$: it transforms between the small component $\chi_k$ of the Dirac spinor and $\varphi_k$, such that the contributions of $\chi_k$ to the electronic eigenstate, as encoded in the (unitless) vector ${\vec b_k=(B_{1k},B_{2k},...,B_{Mk})}$, can be merged with those of $\varphi_k$ with the help of $\tilde{S}$. Upon normalizing each of the so-obtained vectors ${(C_{1k},C_{2k},...,C_{Mk})}$, we obtain the Schr\"odinger-like orthonormal 1pEx-basis $\{\psi_k\}$ and the corresponding interaction tensor elements in Eq.~(\ref{RelTabcd}).

Since spin-dependent contributions are not included in the matrix $U$ of Eq.~(\ref{X2C_U}), the sf-X2C employed here does not recover the (exactly known) hydrogenic eigenenergies of the one-particle Dirac equation \cite{Bethe1957}. Typically, however, the deviations are small: for the example of nuclear charge ${Z=20}$ the lowest energy levels are $-201.077\,\mbox{Ha}$ (from four-component Dirac theory), $-201.055\,\mbox{Ha}$ (from sf-X2C), and $-200\,\mbox{Ha}$ (from nonrelativistic quantum mechanics). Hence, in this case, sf-X2C produces about $98\%$ of the relativistic corrections of the exact Dirac theory.

\begin{center}
\begin{table}[ht]
\setlength\extrarowheight{0.4em}
\setlength\tabcolsep{0.33em}
\begin{tabular}{c|cc|cccc|c|r}
\parbox{3em}{\centering atomic\\[-0.2em] system\\[0.2em]} & $Z$ & $N$ & 1pEx-DFT & HF & \parbox{5em}{\centering 1pEx-DFT\\[-0.2em](sf-X2C)} & \parbox{5em}{\centering HF\\[-0.2em](sf-X2C)} & \parbox{5em}{\centering statistical\\[-0.2em] atom\\[0.2em]} & \parbox{5em}{\centering reference\\[-0.2em] value\\[0.2em]}\\
\hline\hline
H$^-$		&  1 &  2 & 0.48187 	& 0.48793	& 0.48188 	& 0.44883   & --- 		& 0.527731(3)\\
He 			&  2 &  2 & 2.83474 	& 2.85583 	& 2.83489 	& 2.85583 	& 2.73111 	& 2.90339 \\
C$^{4+}$	&  6 &  2 & 32.3146 	& 32.3565  	& 32.3305 	& 32.3735 	& --- 		& 32.416 \\
Ar$^{16+}$	& 18 &  2 & 312.809 	& 312.828  	& 314.166 	& 314.143 	& --- 		& 314.092 \\
Be 			&  4 &  4 & 14.5096 	& 14.5617  	& 14.5129 	& 14.5748 	& 14.2453	& 14.6684 \\
C$_{L=7}$ 	&  6 &  6 & 37.3007 	& 37.5646  	& 37.3198 	& 37.6109 	& 37.6356	& 37.8558 \\
C$_{L=15}$	&  6 &  6 & 37.5097 	& 37.5646  	& 37.5268 	& 37.6109 	& 37.6356 	& 37.8558 \\
C$_{L=31}$ 	&  6 &  6 & 37.5197 	& 37.5646  	& 37.5368 	& 37.6109 	& 37.6356 	& 37.8558 \\
Xe$^{48+}$ 	& 54 &  6 & 4197.74 	& 4199.22  	& 4348.14  	& ---		& --- 		& 4379.7${\pm0.4}$\\
O			&  8 &  8 & 74.1110 	& 74.5883  	& 74.1259 	& 74.7173 	& 75.0362 	& 75.1098 \\
Ne 			& 10 & 10 & 126.064 	& 128.353  	& 126.140 	& 128.625 	& 128.149 	& 129.053 \\
Xe$^{44+}$	& 54 & 10 & 5361.70		& 5373.85  	& 5538.51 	& ---		& --- 		& 5558.4${\pm1.0}$
\end{tabular}
\caption{\label{BindingEnergiesTable}The binding energies (in Hartree) as displayed in Fig.~\ref{RelDev_1pEx_HF_NIST} relative to the (nonrelativistic) HF binding energies. The nonrelativistic HF binding energy for H$^-$ is reported in Ref.~\cite{King2018}. We used the quantum chemistry package PSI4 \cite{Verma2015,Smith2020} to compute all other HF energies displayed in Fig.~\ref{RelDev_1pEx_HF_NIST}, with default PSI4 settings, except: unrestricted HF and `dgauss-dzvp-autoabs-decon' basis set for nonrelativistic HF; `cc-pvdz-dk' basis set for relativistic HF; and we omitted the relativistic HF simulations of the xenon ions for lack of a built-in basis set. The reference value for H$^-$ is the experimental binding energy given in Ref.~\cite{Lykke1991}. All other reference values are taken from Ref.~\cite{NIST2021articleEntry}; where uncertainties are omitted, they are negligible for the number of digits shown.}
\end{table}
\end{center}

\FloatBarrier

\bibliographystyle{bibtexPostdoc}
\setlength{\bibsep}{0.3\baselineskip}
\bibliography{myPostDocbib}

\begin{thebibliography}{10}

\bibitem{Becke2014}
A.~D. Becke, \emph{Perspective: Fifty years of density-functional theory in
  chemical physics}, J. Chem. Phys. \textbf{140}, 18A301 (2014).

\bibitem{Hasnip2014}
P.~J. Hasnip, K.~Refson, M.~I.~J. Probert, J.~R. Yates, S.~J. Clark, and C.~J.
  Pickard, \emph{Density functional theory in the solid state}, Phil. Trans. R.
  Soc. A \textbf{372}, 20130270 (2014).

\bibitem{Okun2023DFMPS}
P.~Okun and K.~Burke, \emph{Semiclassics: The hidden theory behind the success
  of DFT}, arXiv:2106.07839, pp. 179--249 in: Density Functionals for
  Many-Particle Systems: Mathematical Theory and Physical Applications of
  Effective Equations; B.-G. Englert, H. Siedentop, and M.-I. Trappe (eds.);
  Lecture Notes Series, IMS, World Scientific, Singapore  (2023).

\bibitem{Henderson1981}
G.~A. Henderson, \emph{Variational theorems for the single-particle probability
  density and density matrix in momentum space}, Phys. Rev. A \textbf{23}, 19
  (1981).

\bibitem{Cinal1993}
M.~Cinal and B.-G. Englert, \emph{Energy functionals in momentum space:
  Exchange energy, quantum corrections, and the Kohn-Sham scheme}, Phys. Rev. A
  \textbf{48}, 1893 (1993).

\bibitem{Sakurai1995}
Y.~Sakurai, Y.~Tanaka, A.~Bansil, S.~Kaprzyk, A.~T. Stewart, Y.~Nagashima,
  T.~Hyodo, S.~Nanao, H.~Kawata, and N.~Shiotani, \emph{High-Resolution Compton
  Scattering Study of Li: Asphericity of the Fermi Surface and Electron
  Correlation Effects}, Phys. Rev. Lett. \textbf{74}, 2252 (1995).

\bibitem{Hueck2018}
K.~Hueck, N.~Luick, L.~Sobirey, J.~Siegl, T.~Lompe, and H.~Moritz,
  \emph{Two-Dimensional Homogeneous Fermi Gases}, Phys. Rev. Lett.
  \textbf{120}, 060402 (2018).

\bibitem{Levy1979}
M.~Levy, \emph{Universal variational functionals of electron densities,
  first-order density matrices, and natural spin-orbitals and solution of the
  v-representability problem}, Proc. Natl. Acad. Sci. \textbf{76}, 6062 (1979).

\bibitem{Levy1995}
M.~Levy and A.~G\"orling, \emph{Correlation-energy density-functional formulas
  from correlating first-order density matrices}, Phys. Rev. A \textbf{52},
  R1808 (1995).

\bibitem{Savin1995}
A.~Savin, \emph{Expression of the exact electron-correlation-energy density
  functional in terms of first-order density matrices}, Phys. Rev. A
  \textbf{52}, R1805 (1995).

\bibitem{Goedecker1998}
S.~Goedecker and C.~J. Umrigar, \emph{Natural Orbital Functional for the
  Many-Electron Problem}, Phys. Rev. Lett. \textbf{81}, 866 (1998).

\bibitem{Cioslowski2000}
J.~Cios{\l}owski~(ed.), \emph{Many-Electron Densities and Reduced Density
  Matrices}, Springer, New York  (2000).

\bibitem{Giesbertz2019}
K.~J.~H. Giesbertz and M.~Ruggenthaler, \emph{One-body reduced density-matrix
  functional theory in finite basis sets at elevated temperatures}, Phys. Rep.
  \textbf{806}, 1 (2019).

\bibitem{Englert1982}
B.-G. Englert and J.~Schwinger, \emph{Thomas--Fermi revisited: The outer
  regions of the atom}, Phys. Rev. A \textbf{26}, 2322 (1982).

\bibitem{Trappe2016}
M.-I. Trappe, Y.~L. Len, H.~K. Ng, C.~A. M\"uller, and B.-G. Englert,
  \emph{Leading gradient correction to the kinetic energy for two-dimensional
  fermion gases}, Phys. Rev. A \textbf{93}, 042510 (2016).

\bibitem{Trappe2017}
M.~I. Trappe, Y.~L. Len, H.~K. Ng, and B.~G. Englert, \emph{Airy-averaged
  gradient corrections for two-dimensional fermion gases}, Ann. Phys. (N. Y.)
  \textbf{385}, 136 (2017).

\bibitem{Chau2018}
T.~T. Chau, J.~H. Hue, M.-I. Trappe, and B.-G. Englert, \emph{Systematic
  corrections to the Thomas--Fermi approximation without a gradient expansion},
  New J. Phys. \textbf{20}, 073003 (2018).

\bibitem{Englert2019articleEntry}
B.-G. Englert, \emph{Julian Schwinger and the Semiclassical Atom},
  arXiv:1907.04751, Chapter 17, pp. 261-269 in: Proceedings of the Julian
  Schwinger Centennial Conference; B.-G. Englert (ed.); World Scientific
  (2019).

\bibitem{Lieb1983}
E.~H. Lieb, \emph{Density functionals for coulomb systems}, Int. J. Quantum
  Chem. \textbf{24}, 243 (1983).

\bibitem{Englert2023DFMPS}
B.-G. Englert, J.~H. Hue, Z.~C. Huang, M.~M. Paraniak, and M.-I. Trappe,
  \emph{Energy functionals of single-particle densities: A unified view},
  arXiv:2206.10097, pp. 287--308 in: Density Functionals for Many-Particle
  Systems: Mathematical Theory and Physical Applications of Effective
  Equations; B.-G. Englert, H. Siedentop, and M.-I. Trappe (eds.); Lecture
  Notes Series, IMS, World Scientific, Singapore  (2023).

\bibitem{Cioslowski2023}
J.~Cioslowski, B.-G. Englert, M.-I. Trappe, and J.~H. Hue, \emph{Contactium: A
  strongly correlated model system}, J. Chem. Phys. \textbf{158}, 184110
  (2023).

\bibitem{Trappe2021b}
M.-I. Trappe, P.~T. Grochowski, J.~H. Hue, T.~Karpiuk, and
  K.~Rz\k{a}$\dot{\mbox{z}}$ewski, \emph{Phase Transitions of Repulsive
  Two-Component Fermi Gases in Two Dimensions}, New J. Phys. \textbf{23},
  103042 (2021).

\bibitem{Dirac1930}
P.~A.~M. Dirac, \emph{Note on the exchange phenomena in the Thomas Atom}, Math.
  Proc. Camb. Philos. Soc. \textbf{26}, 376 (1930).

\bibitem{Hartree1928}
D.~R. Hartree, \emph{The Wave Mechanics of an Atom with a Non-Coulomb Central
  Field. Part I. Theory and Methods}, Math. Proc. Camb. Philos. Soc.
  \textbf{24}, 89 (1928).

\bibitem{Slater1930}
J.~C. Slater, \emph{Note on Hartree's Method}, Phys. Rev. \textbf{35}, 210
  (1930).

\bibitem{Fock1930}
V.~Fock, \emph{N{\"a}herungsmethode zur L{\"o}sung des quantenmechanischen
  Mehrk{\"o}rperproblems}, Z. Phys. \textbf{61}, 126 (1930).

\bibitem{Kennedy1995}
J.~Kennedy and R.~Eberhart, \emph{Particle swarm optimization}, pp. 1942--1948
  in: Proceedings of ICNN'95 - International Conference on Neural Networks;
  IEEE  (1995).

\bibitem{Bonyadi2017}
R.~Bonyadi, M and Z.~Michalewicz, \emph{Particle Swarm Optimization for Single
  Objective Continuous Space Problems: A Review}, Evol. Comput. \textbf{25}, 1
  (2017).

\bibitem{Tang2021}
J.~Tang, G.~Liu, and Q.~Pan, \emph{A review on representative swarm
  intelligence algorithms for solving optimization problems: Applications and
  trends}, IEEE/CAA J. Autom. Sin. \textbf{8}, 1627 (2021).

\bibitem{Slowik2020}
A.~Slowik and H.~Kwasnicka, \emph{Evolutionary algorithms and their
  applications to engineering problems}, Neural. Comput. Appl. \textbf{32},
  12363 (2020).

\bibitem{ParaniakPhdThesis2022}
M.~M. Paraniak, \emph{(A) Reduced Density Matrix Generation Algorithms in
  Single-Particle-Exact Density Functional Theory, (B) Quantum Dynamical
  Simulation of a Transversal Stern-Gerlach Interferometer, (C) Open Quantum
  System Process Tomography}, Ph.D. thesis, National University of Singapore,
  Singapore (2022).

\bibitem{deSouza2010}
S.~Xavier-de Souza, J.~A.~K. Suykens, J.~Vandewalle, and D.~Bolle,
  \emph{Coupled Simulated Annealing}, IEEE Transactions on Systems, Man, and
  Cybernetics, Part B (Cybernetics) \textbf{40}, 320 (2010).

\bibitem{Kutzelnigg2005}
W.~Kutzelnigg and W.~Liu, \emph{Quasirelativistic theory equivalent to fully
  relativistic theory}, J. Chem. Phys. \textbf{123}, 241102 (2005).

\bibitem{Liu2009}
W.~Liu and D.~Peng, \emph{Exact two-component Hamiltonians revisited}, J. Chem.
  Phys. \textbf{131}, 031104 (2009).

\bibitem{Cheng2011}
L.~Cheng and J.~Gauss, \emph{Analytic energy gradients for the spin-free exact
  two-component theory using an exact block diagonalization for the
  one-electron Dirac Hamiltonian}, J. Chem. Phys. \textbf{135}, 084114 (2011).

\bibitem{Cunha2022}
L.~A. Cunha, D.~Hait, R.~Kang, Y.~Mao, and M.~Head-Gordon, \emph{Relativistic
  Orbital-Optimized Density Functional Theory for Accurate Core-Level
  Spectroscopy}, J. Phys. Chem. Lett. \textbf{13}, 3438 (2022).

\bibitem{Trappe2019}
M.-I. Trappe, D.~Y.~H. Ho, and S.~Adam, \emph{First-principles quantum
  corrections for carrier correlations in double-layer two-dimensional
  heterostructures}, Phys. Rev. B \textbf{99}, 235415 (2019).

\bibitem{Trappe2023DFMPS}
M.-I. Trappe, J.~H. Hue, and B.-G. Englert, \emph{Density-potential functional
  theory for fermions in one dimension}, arXiv:2106.07839, pp. 251--267 in:
  Density Functionals for Many-Particle Systems: Mathematical Theory and
  Physical Applications of Effective Equations; B.-G. Englert, H. Siedentop,
  and M.-I. Trappe (eds.); Lecture Notes Series, IMS, World Scientific,
  Singapore  (2023).

\bibitem{TrappeWittManzhosXXXX}
M.-I. Trappe, C.~Witt, and S.~Manzhos, \emph{Atoms, dimers, and nanoparticles
  from orbital-free density-potential functional theory} (in preparation).

\bibitem{Trappe2023NatComm}
M.-I. Trappe and R.~A. Chisholm, \emph{A density functional theory for ecology
  across scales}, Nat. Commun. \textbf{14}, 1089 (2023).

\bibitem{Saunders1973}
V.~R. Saunders and I.~H. Hillier, \emph{A "Level-Shifting" method for
  converging closed shell Hartree-Fock wave functions}, Int. J. Quantum Chem.
  \textbf{7}, 699 (1973).

\bibitem{Busch+3:98}
T.~Busch, B.-G. Englert, K.~Rz\k{a}\.{z}ewski, and M.~Wilkens, \emph{Two Cold
  Atoms in a Harmonic Trap}, Found. Phys. \textbf{28}, 549 (1998).

\bibitem{Viana-Gomes+1:11}
J.~Viana-Gomes and N.~M.~R. Peres, \emph{{Solution of the quantum harmonic
  oscillator plus a delta-function potential at the origin: The oddness of its
  even-parity solutions}}, Eur. J. Phys. \textbf{32}, 1377 (2011).

\bibitem{BenavidesRiveros2014}
C.~L. Benavides-Riveros, I.~V. Toranzo, and J.~S. Dehesa, \emph{Entanglement in
  N-harmonium: bosons and fermions}, J. Phys. B: At. Mol. Opt. Phys.
  \textbf{47}, 195503 (2014).

\bibitem{NIST2021articleEntry}
A.~Kramida, Y.~Ralchenko, J.~Reader, and $\mbox{NIST ASD Team}$, \emph{NIST
  Atomic Spectra Database (ver. 5.9), [Online]}, available:
  https://physics.nist.gov/asd, National Institute of Standards and Technology,
  Gaithersburg, MD  (2021).

\bibitem{Fischer2016}
C.~F. Fischer, M.~Godefroid, T.~Brage, P.~J\"onsson, and G.~Gaigalas,
  \emph{Advanced multiconfiguration methods for complex atoms: I. Energies and
  wave functions}, J. Phys. B: At. Mol. Opt. Phys. \textbf{49}, 182004 (2016).

\bibitem{Bethe1957}
H.~Bethe and E.~Salpeter, \emph{Quantum Mechanics of One- and Two-Electron
  Systems}, Springer, Berlin, Heidelberg  (1957).

\bibitem{Scott1952}
J.~Scott, \emph{LXXXII. The binding energy of the Thomas-Fermi Atom}, Lond.
  Edinb. Dublin philos. mag. j. sci. \textbf{43}, 859 (1952).

\bibitem{Berge1988articleEntry}
B.-G. Englert, \emph{Lecture Notes in Physics: Semiclassical Theory of Atoms},
  Springer, Berlin, Heidelberg  (1988).

\bibitem{Lykke1991}
K.~R. Lykke, K.~K. Murray, and W.~C. Lineberger, \emph{Threshold
  photodetachment of $H^-$}, Phys. Rev. A \textbf{43}, 6104 (1991).

\bibitem{Kingma2017}
D.~P. Kingma and J.~Ba, \emph{Adam: A Method for Stochastic Optimization},
  arXiv:1412.6980v9 [cs.LG]  (2017).

\bibitem{Lehman2008}
J.~Lehman and K.~O. Stanley, \emph{Exploiting Open-Endedness to Solve Problems
  Through the Search for Novelty}, in: Proceedings of the Eleventh
  International Conference on Artificial Life (ALIFE XI), MIT Press, Cambridge,
  MA  (2008).

\bibitem{LaTorre2015}
A.~LaTorre, S.~Muelas, and J.-M. Pe$\tilde{\mbox{n}}$a, \emph{A comprehensive
  comparison of large scale global optimizers}, Inf. Sci. \textbf{316} (2015).

\bibitem{Glorieux2015}
E.~Glorieux, B.~Svensson, F.~Danielsson, and B.~Lennartson, \emph{Improved
  Constructive Cooperative Coevolutionary Differential Evolution for
  Large-Scale Optimisation}, pp.~1703--1710 in: IEEE Symposium Series on
  Computational Intelligence  (2015).

\bibitem{Sun2018}
Y.~Sun, M.~Kirley, and S.~K. Halgamuge, \emph{A Recursive Decomposition Method
  for Large Scale Continuous Optimization}, IEEE Trans. Evol. Comput.
  \textbf{22}, 647 (2018).

\bibitem{Cioslowski2019}
J.~Cioslowski, Z.~E. Mihalka, and A.~Szabados, \emph{Bilinear constraints upon
  the correlation contribution to the electron--electron repulsion energy as a
  functional of the one-electron reduced density matrix}, J. Chem. Theory
  Comput. \textbf{15}, 4862 (2019).

\bibitem{Liu2014}
Q.~Liu, \emph{Order-2 Stability Analysis of Particle Swarm Optimization}, Evol.
  Comput. \textbf{23}, 187 (2014).

\bibitem{Nobile2018}
M.~S. Nobile., P.~Cazzaniga, D.~Besozzi, R.~Colombo, G.~Mauri, and G.~Pasi,
  \emph{Fuzzy Self-Tuning PSO: A settings-free algorithm for global
  optimization}, Swarm Evol. Comput. \textbf{39}, 70 (2018).

\bibitem{Hutter2006}
M.~Hutter and S.~Legg, \emph{Fitness uniform optimization}, IEEE Trans. Evol.
  Comput. \textbf{10}, 568 (2006).

\bibitem{Park2010}
T.~Park and K.~R. Ryu, \emph{A dual-population genetic algorithm for adaptive
  diversity control}, IEEE Trans. Evol. Comput. \textbf{14}, 865 (2010).

\bibitem{Na+1:17}
M.~Na and F.~Marsuglio, \emph{{Two and three particles interacting in a
  one-dimensional trap}}, Am. J. Phys. \textbf{85}, 769 (2017).

\bibitem{Condon1935}
E.~U. Condon and G.~H. Shortley, \emph{The Theory of Atomic Spectra}, Cambridge
  University Press, New York  (1935).

\bibitem{NISTHandbook}
F.~W.~J. Olver, D.~W. Lozier, R.~F. Boisvert, and C.~W. Clark, \emph{NIST
  Handbook of Mathematical Functions}, Cambridge University Press, New York
  (2010).

\bibitem{Joshi1971}
B.~D. Joshi, \emph{Generalized Form of Atomic Two Electron Integrals over
  Hydrogenic Functions}, J. Comp. Phys. \textbf{8}, 300 (1971).

\bibitem{Butler1971}
P.~H. Butler, P.~E.~H. Minchin, and B.~G. Wybourne, \emph{Tables of Hydrogenic
  Slater Radial Integrals}, At. Data Nucl. Data Tables \textbf{3}, 153 (1971).

\bibitem{Butler1971Erratum}
P.~H. Butler, P.~E.~H. Minchin, and B.~G. Wybourne, \emph{Tables of Hydrogenic
  Slater Radial Integrals}, At. Data Nucl. Data Tables \textbf{27}, 145 (1982),
  erratum: At. Data Nucl. Data Tables \textbf{3}, 153 (1971).

\bibitem{King2018}
A.~W. King, A.~L. Baskerville, and H.~Cox, \emph{Hartree--Fock implementation
  using a Laguerre-based wave function for the ground state and correlation
  energies of two-electron atoms}, Philos. Trans. R. Soc. A \textbf{376},
  20170153 (2018).

\bibitem{Verma2015}
P.~Verma, W.~D. Derricotte, and F.~A. Evangelista, \emph{Predicting Near Edge
  X-ray Absorption Spectra with the Spin-Free Exact-Two-Component Hamiltonian
  and Orthogonality Constrained Density Functional Theory}, J. Chem. Theory
  Comput. \textbf{12}, 144 (2016).

\bibitem{Smith2020}
D.~G.~A. Smith~et al., \emph{PSI4 1.4: Open-source software for high-throughput
  quantum chemistry}, J. Chem. Phys. \textbf{152}, 184108 (2020).

\end{thebibliography}





\end{document}